\documentclass[12pt]{article}
\usepackage{amsmath}
\usepackage{amsbsy}
\usepackage{graphicx}
\usepackage{enumerate}
\usepackage{natbib}
\usepackage{url} 
\usepackage{binomexp}
\usepackage{mathtools}
\usepackage{amsfonts}
\usepackage{mathrsfs}
\RequirePackage{subcaption}
\usepackage{bbm}
\usepackage{dsfont}
\usepackage{longtable}
\usepackage{booktabs}
\usepackage{yhmath}
\usepackage{bm}
\usepackage{amssymb}
\usepackage[nodisplayskipstretch]{setspace}
\usepackage{tikz}
\usepackage{float}
\usepackage{color}
\usepackage{stmaryrd}
\usepackage{paralist}
\usepackage{multirow}
\usepackage{threeparttable}
\usepackage{soul}
\usepackage{algorithm,algpseudocode}

\makeatletter
\newcommand{\algmargin}{\the\ALG@thistlm}
\makeatother
\newlength{\whilewidth}
\algdef{SE}[parWHILE]{parWhile}{EndparWhile}[1]
  {\parbox[t]{\dimexpr\linewidth-\algmargin}{%
     \hangindent\whilewidth\strut\algorithmicwhile\ #1\ \algorithmicdo\strut}}{\algorithmicend\ \algorithmicwhile}%
\algnewcommand{\parState}[1]{\State%
  \parbox[t]{\dimexpr\linewidth-\algmargin}{\strut #1\strut}}

\newcommand{\bfo}{\bm{o}}

\newcommand{\bfu}{\bm{u}}
\newcommand{\bfv}{\bm{v}}

\newcommand{\bfbeta}{\bm{\beta}}

\newcommand{\bflambda}{\bm{\lambda}}
\newcommand{\bfvartheta}{\bm{\vartheta}}

\newcommand{\bfpsi}{\bm{\psi}}

\newcommand{\bfSigma}{\bm{\Sigma}}

\newcommand{\np}{{\mathrm{np}}}

\newcommand{\mr}{{\mathrm{mr}}}

\newcommand{\eff}{{\mathrm{eff}}}

\newcommand{\Var}{{\mathbb{V}\mathrm{ar}}}

\newcommand{\si}{{\mathrm{si}}}
\newcommand{\pr}{{\mathrm{pr}}}
\newcommand{\dml}{{\mathrm{dml}}}
\newcommand{\est}{{\mathrm{est}}}

\newcommand{\bfA}{\bm{A}}
\newcommand{\bfB}{\bm{B}}
\newcommand{\bfC}{\bm{C}}

\newcommand{\bfO}{\bm{O}}

\newcommand{\bfU}{\bm{U}}
\newcommand{\bfV}{\bm{V}}

\newcommand{\bfX}{\bm{X}}
\newcommand{\bfY}{\bm{Y}}

\newcommand{\bbE}{\mathbb{E}}

\newcommand{\bbN}{\mathbb{N}}
\newcommand{\bbP}{\mathbb{P}}
\newcommand{\bbR}{\mathbb{R}}

\newcommand{\bbone}{\mathbbm{1}}

\newcommand{\calA}{\mathcal{A}}

\newcommand{\calD}{\mathcal{D}}
\newcommand{\calE}{\mathcal{E}}

\newcommand{\calI}{\mathcal{I}}

\newcommand{\calL}{\mathcal{L}}

\newcommand{\calN}{\mathcal{N}}
\newcommand{\calO}{\mathcal{O}}
\newcommand{\calP}{\mathcal{P}}

\newcommand{\scrW}{\mathscr{W}}
\newcommand{\scrY}{\mathscr{Y}}

\newcommand{\frakw}{\mathfrak{w}}

\newcommand{\frakI}{\mathfrak{I}}

\usepackage[colorlinks=true,linkcolor=blue,citecolor=blue]{hyperref}

\newtheorem{theorem}{Theorem}
\newtheorem{assumption}{Assumption}

\newtheorem{remark}{Remark}
\newtheorem{example}{Example}

\newcommand{\blind}{0}

\allowdisplaybreaks

\usetikzlibrary{shapes,decorations,arrows,calc,arrows.meta,fit,positioning}
\tikzset{
     -Latex,auto,node distance =1 cm and 1 cm,semithick,
     state/.style ={ellipse, draw, minimum width = 0.7 cm},
     point/.style = {circle, draw, inner sep=0.04cm,fill,node contents={}},
    bidirected/.style={Latex-Latex,dashed},
     el/.style = {inner sep=2pt, align=left, sloped}
 }

\begin{document}

\def\spacingset#1{\renewcommand{\baselinestretch}%
{#1}\small\normalsize} \spacingset{1}


\if0\blind
{
    \begin{center}
        {\Large \bf Optimal estimation of generalized causal effects in cluster-randomized trials with multiple outcomes}\\[0.5cm]
        {\large Xinyuan Chen$^{1,\ast}$ and Fan Li$^2$\\[0.2cm]
        $^1$Department of Mathematics and Statistics,\\ Mississippi State University, MS, USA\\
        $^2$Department of Biostatistics, Yale School of Public Health, CT, USA\\
        ${}^\ast$xchen@math.msstate.edu}
    \end{center}
} \fi

\if1\blind
{
  \bigskip
  \bigskip
  \bigskip
  \begin{center}
    {\Large \bf Optimal estimation of generalized causal effects in cluster-randomized trials with multiple outcomes}
\end{center}
  \medskip
} \fi

\begin{abstract}
    Cluster-randomized trials (CRTs) are used to evaluate group-level interventions and can collect multiple outcomes capturing complementary dimensions of benefit and risk. Investigators typically seek a single global summary of treatment effect, yet existing methods largely focus on single-outcome estimands or rely on model-based procedures with unclear causal interpretation or limited robustness. We develop a unified potential outcomes framework for generalized treatment effects with multiple outcomes in CRTs, accommodating both non-prioritized and prioritized outcome settings. The proposed cluster-pair and individual-pair causal estimands are defined through pairwise contrast functions and explicitly account for informative cluster size. We establish nonparametric estimation via weighted clustered U-statistics and derive efficient influence functions to construct covariate-adjusted estimators that integrate debiased machine learning with U-statistics. The resulting estimators are consistent and asymptotically normal, attain semiparametric efficiency bounds under mild regularity conditions, and admit analytically tractable variance estimators that are consistent under cross-fitting. Simulations and an application to a CRT for chronic pain management illustrate the practical utility of our proposed methods.
\end{abstract}

\noindent%

\spacingset{1.75} 

\section{Introduction} \label{sec:intro}

Cluster randomized trials (CRTs) are a fundamental design for evaluating interventions that are implemented at the group level, such as clinics, schools, or communities, and have become increasingly prevalent in biomedical and social science research. Although treatment is assigned at the cluster level, outcome observations are typically measured on individuals, subject to within‐cluster dependence and unequal cluster sizes that must be accounted for when estimating treatment effects. CRTs have traditionally been analyzed using model‐based approaches \citep{Turner2017review}; however, an emerging literature has focused on estimating average treatment effects defined under the potential outcomes framework. For example, \citet{Su2021} established the asymptotic properties of model‐assisted regression-based estimators for weighted average treatment effects in CRTs. \citet{wang2024model} developed model‐robust and efficient covariate-adjusted estimators for cluster‐average and individual‐average treatment effect estimands based on efficient influence functions. Several other notable contributions include \citet{balzer2019new}, \citet{Bugni2024}, and \citet{li2025model}, which collectively advance model-robust causal inference approaches in CRTs.

{\color{black}In recent research, CRTs increasingly collect multiple outcomes that capture complementary dimensions of benefit and risk, motivating the use of a single summary measure to provide an initial assessment of the overall treatment effect.} Broadly, two types of multiple outcome settings have emerged. The first type concerns non-prioritized outcomes, in which components may be measured on different scales, but none is assigned intrinsic clinical precedence. In this setting, summary measures that map heterogeneous outcomes onto a common probability scale are appealing; for example, through the global treatment effect summary in \citet{Smith2025}. A second type involves prioritized outcomes, in which components are ordered by clinical importance and compared sequentially. This setup underlies hierarchical composite outcomes and generalized pairwise comparison estimands \citep{Pocock2012}. The extant literature on CRTs has focused on single-outcome settings, with only limited attention to multiple outcomes. Exceptions include \citet{Smith2025}, who developed a model-based approach for global treatment effect estimation with multiple non-prioritized outcomes, and \citet{fang2025sample}, who proposed nonparametric testing and sample size methods for multiple prioritized outcomes. 

Despite these developments, several gaps remain in causal inference with multiple outcomes in CRTs. First, well-defined potential outcome estimands for scalar generalized causal effects (GCEs) that explicitly accommodate heterogeneous outcome types and potentially informative cluster sizes are lacking \citep{kahan2023estimands}. {\color{black}\label{par:gce-def}Here, by GCEs we refer to the class of summary metrics constructed from generalized pairwise comparisons of potential outcomes \citep{buyse2010generalized}, including the net benefit (the difference between the probability that a treated unit fares better than a control unit and its converse), the win ratio (the corresponding ratio), and the win odds (which credits ties). These estimands have gained increasing popularity for analysis of independent data, but their extension to CRTs---in particular, how clusters of unequal size should be weighted---remains understudied.} Second, although covariate adjustment is known to improve precision for single outcome estimands in CRTs \citep{balzer2019new,wang2024model}, existing multiple outcome methods generally ignore baseline covariates \citep{Smith2025,fang2025sample}, and it is not obvious how to reintroduce them without perturbing the target estimand. Finally, estimators that combine flexible nuisance modeling with efficiency guarantees under cluster randomization have not been developed for GCEs with multiple outcomes.

Our contributions are hence threefold. First, we develop a unified potential outcomes framework defining GCE estimands for CRTs with multiple outcomes, non-prioritized and prioritized alike, while accounting for informative cluster sizes. We introduce the cluster-pair and individual-pair GCE, which equally weight cluster pairs and individual pairs, respectively; both are defined through a general, possibly nonlinear contrast function and accommodate outcomes of mixed types, and both reduce to the conventional cluster-average and individual-average treatment effects when the contrast is a difference of a single outcome \citep{kahan2023estimands}. Second, we establish nonparametric identification and estimation under cluster randomization, with estimators taking the form of two-sample U-statistics for clustered data \citep{Lee2005}. {\color{black}\label{par:contrib3}Third, we develop the semiparametric efficiency theory for these estimands. We derive the efficient influence functions (EIFs) for each estimand, characterizing the information bounds under informative cluster size, and use the resulting orthogonal estimating equations to build covariate-adjusted U-statistics. Debiased machine learning (DML) \citep{Chernozhukov2018, chen2024principal} then serves as the computational mechanism for attaining those bounds while retaining robustness to misspecification, and the orthogonal structure further yields a tractable analytical variance estimator under clustered sample splitting, which we also prove to be consistent.} To relieve the computational cost of U-statistics, we also propose an estimator based on incomplete U-statistics \citep{blom1976some} and show that it remains asymptotically efficient.

\section{Notation, estimands and assumptions} \label{sec:setting}

We consider a CRT with $m$ clusters, each containing $N_i$ individuals. We define $A_i=a\in\{0,1\}$ as the cluster-level treatment assignment, and $\bfC_i$ as the $p_c$-dimensional vector of baseline cluster-level covariates (e.g., geographical location). For each individual $j$ in cluster $i$, we define $\bfY_{ij}=(Y_{1,ij},\ldots,Y_{Q,ij})^\top$ as the $Q\geq 1$ individual-level outcomes that are of simultaneous interest; here, each component outcome $Y_{q,ij}$ can be continuous, binary, count, ordinal, or categorical and possibly measured on a different scale. We denote by $\bfX_{ij}$ the $p_x$-dimensional vector of individual-level baseline covariates (e.g., demographics and baseline diagnosis). We aim to obtain a single summary measure of the treatment effect across multiple outcomes $\bfY = (Y_1, \ldots, Y_Q)^\top$ that jointly reflect overall benefit or risk. When multiple outcomes are analyzed simultaneously, two distinct settings arise. In the first setting, there is no clear consensus regarding the relative clinical importance of the outcomes; such outcomes are referred to as \emph{non-prioritized}. In the second setting, existing clinical knowledge allows ranking outcomes by clinical importance; in this case, the outcomes are referred to as \emph{prioritized}. {\color{black} A summary table of mathematical notation used throughout is provided in Web Table S1 of the Supplementary Materials.}

Under the Neyman-Rubin potential outcomes framework, we define the potential outcome vector as $\bfY(a)=(Y_1(a),\allowbreak\ldots,Y_Q(a))^\top$, where $Y_q(a)\in\scrY_q$ denotes the $q$th potential outcome under treatment condition $a\in\{0,1\}$. 

\begin{assumption}[Cluster-level SUTVA] \label{asp:sutva}
    Let $\bfY_{ij}(A_i,\allowbreak \bfA_{-i})$ denote the potential outcome vector for individual $j$ in cluster $i$ given assignments across all clusters, and $\bfA_{-i}$ denote the vector of assignments for all other clusters, then (i) $\bfY_{ij}(A_i,\bfA_{-i}) = \bfY_{ij}(A_i,\bfA_{-i}^*)$ for all $\bfA_{-i}\neq \bfA_{-i}^*$; and (ii) $\bfY_{ij}(A_i,\bfA_{-i}) = \bfY_{ij}(A_i^*,\bfA_{-i}^*)$ if $A_i=A_i^*$.
\end{assumption}

Assumption \ref{asp:sutva} requires that individual-level potential outcomes in cluster $i$ only depend on the treatment assignment for cluster $i$, hence ruling out between-cluster interference and ensuring that $\bfY_{ij}(a)$ is well-defined. As a result, $\bfY_{ij}=A_i\bfY_{ij}(1)+(1-A_i) \bfY_{ij}(0)$. To formulate our causal estimands, we define an arbitrary, possibly nonlinear contrast function between two given potential outcome vectors under different conditions, $a$ and $1-a$, as 
\begin{equation} \label{eq:contrast}
    \text{contrast function}: w\{\bfY(a),\bfY(1-a)\},~~~w:\pmb\scrY\times\pmb\scrY\mapsto\scrW\subseteq\bbR,
\end{equation}
where $\pmb\scrY=\times_{q=1}^Q\scrY_q$, is the Cartesian product of all individual supports. The contrast function represents a rule that maps a multivariate comparison to a single numerical value.

Structurally, the contrast function $w$ in \eqref{eq:contrast} can be constructed via two strategies: (i) dimension-wise comparison, which aggregates the results from outcome-specific comparisons, or (ii) joint comparison, which evaluates multiple outcomes directly. Under the dimension-wise approach, $w(\bfu,\bfv) = \calA\{w_1(u_1, v_1), \ldots, w_Q(u_Q, v_Q)\}$, where $w_q(u_q,v_q):\scrY_q\times\scrY_q\mapsto\scrW_q\subseteq\bbR$ is the contrast function applied to the $q$-th outcome, and $\calA: \times_{q=1}^Q \scrW_q \mapsto \scrW$ is an aggregation operator. For example, $\calA = \sum_{q=1}^Q\omega_q w_q$ can be defined with pre-specified outcome-specific weights $\omega_q\geq0$ satisfying $\sum_{q=1}^Q\omega_q=1$ \citep{Smith2025}. Under the joint approach, one directly defines a rule $w$ on $\pmb\scrY\times\pmb\scrY$, e.g., with a prioritized comparison rule $w(\bfu,\bfv)=\bbone(\bfu\succ \bfv)$ \citep{bebu2016large} or the non-prioritized Pareto partial order $w(\bfu,\bfv)=\bbone(\bfu\preceq_P \bfv)$. These two strategies can also be combined, e.g., by applying joint comparisons within pre-specified outcome groups and then aggregating the results dimension-wise across groups. We give several examples below.

{\color{black}
\begin{example}[\emph{Non-prioritized, dimension-wise comparison}]\label{ex:non-prioritized} 
    The non-prioritized setting involves multiple outcomes that do not admit a hierarchical order of importance; the $Q$ outcomes are compared simultaneously. Consider the Pain Program for Active Coping and Training (PPACT) chronic pain application in Section \ref{sec:data}, where investigators evaluate the PEGS (Pain, Enjoyment of life, General activity, Sleep) pain impact scale ($Y_1$) and the Roland-Morris Disability Questionnaire (RMDQ) disability score ($Y_2$). If neither is deemed more important, one may specify an outcome-specific contrast based on clinically meaningful margins. For example, using the Heaviside step function with a threshold $c_q$, $w_q(u_q, v_q) = \bbone(u_q < v_q - c_q) + 0.5\bbone(|u_q - v_q| \le c_q)$. The overall contrast is then $w(u, v) = \omega_1 w_1(u_1, v_1) + \omega_2 w_2(u_2, v_2)$, where the weights $\omega_q$ quantify the pre-specified relative contribution of each outcome; by default one can specify $\omega_1=\omega_2=1/2$.
\end{example}

\begin{example}[\emph{Prioritized, joint comparison}]\label{ex:prioritized}
    With prioritized outcomes, a consensus-based hierarchical ordering is imposed on $(Y_1, \ldots, Y_Q)$. In the PPACT trial, if reducing pain ($Y_1$) is prioritized over disability ($Y_2$), the hierarchy is formally represented through a joint comparison $w(u,v) = \bbone(u \succ v)$, where $\succ$ denotes a user-specified comparison rule. If the comparison on $Y_1$ is inconclusive (a tie within a clinical equivalence margin), the contrast proceeds to $Y_2$. Specifically, $w(u,v) = \bbone(u_1 < v_1 - c_1) + \bbone(|u_1 - v_1| \le c_1)\bbone(u_2 \succ v_2)$.
\end{example}

\begin{example}[\emph{Non-prioritized, joint comparison}]\label{ex:pareto}
    The Pareto partial order defines a relationship where one intervention is deemed preferable if it is strictly superior in at least one outcome while being no worse in any others. In the PPACT trial involving pain severity ($Y_1$), disability ($Y_2$), and patient satisfaction ($Y_3$), with satisfaction reverse-coded, the contrast function becomes $w(u,v) = \bbone(u \le_P v)$, where $\bbone(u \le_P v) = 1$ if $u_q \le v_q$ for all $q \in \{1,2,3\}$ and $u_q < v_q$ for some $q$. 
\end{example}
}

{\color{black}\label{par:wguide}The specification of $w(\cdot)$ requires clinical justification, since it defines the target estimand. Accordingly, the weights $\omega_q$, equivalence margins $c_q$, and prioritization hierarchy are substantive choices rather than tuning parameters to be selected from the data. They should be prespecified using clinical evidence, including published minimal clinically important differences where applicable, together with input from clinicians and patients. Qualitative research, such as interviews or focus groups, can help identify relevant outcomes and understand preferences regarding their relative importance, while formal consensus methods, such as a Delphi process, can help establish agreement on the weights, equivalence margins, or prioritization hierarchy; see also \citet{fang2025while} for a related discussion. Because reasonable experts may disagree, we also recommend accompanying the primary analysis with sensitivity analyses across plausible alternative kernels to assess the stability of conclusions, as illustrated in Section \ref{subsec:sensitivity}. Although pairwise estimands may appear more abstract than mean differences, they address a concrete question: ``\textit{under the specified comparison rule, what is the expected advantage of a randomly selected unit under treatment over an independently selected unit under control?}'' They may be preferable to mean contrasts when outcomes have mixed types that preclude meaningful arithmetic averaging, or when clinical interest centers on relative ordering rather than magnitudes, particularly for highly skewed outcomes. Interpretation may become more challenging as $Q$ increases, especially when the comparison rule lacks a clinically justified hierarchy or weighting scheme that connects the resulting scalar to a coherent clinical question.}

Given a contrast function, we define the cluster-pair and individual-pair GCE (cp-GCE and ip-GCE) estimands. Let $N_1$ and $N_2$ denote two independent draws from the marginal cluster size distribution $\calP(N)$, and $\{\mathbf Y_1(a),\mathbf Y_1(1-a)\}$ and $\{\mathbf Y_2(a),\mathbf Y_2(1-a)\}$ the corresponding potential outcomes independently sampled from the conditional distribution $\calP(\mathbf Y(a),\mathbf Y(1-a)|N)$, where $\mathbf Y_i(a)=\{\bfY_{ij}(a),j=1,\ldots,N_i\}$ for $a\in\{0,1\}$ and $i=1,2$. The cp-GCE and ip-GCE are respectively defined as
\begin{align} 
    \lambda_{C,a} =& \bbE\left[\frac{\sum_{j=1}^{N_1}\sum_{l=1}^{N_2} w\{\bfY_{1j}(a),\bfY_{2l}(1-a)\}}{N_1N_2}\right], \label{eq:estimand-C}\displaybreak[0]\\
    \lambda_{I,a} =& \frac{\bbE\left[\sum_{j=1}^{N_1}\sum_{l=1}^{N_2} w\{\bfY_{1j}(a),\bfY_{2l}(1-a)\}\right]}{\bbE(N_1N_2)}. \label{eq:estimand-I}
\end{align}

\begin{figure}[ht!]
    \vspace*{0in}
    \centering
    \includegraphics[width = 0.9\textwidth]{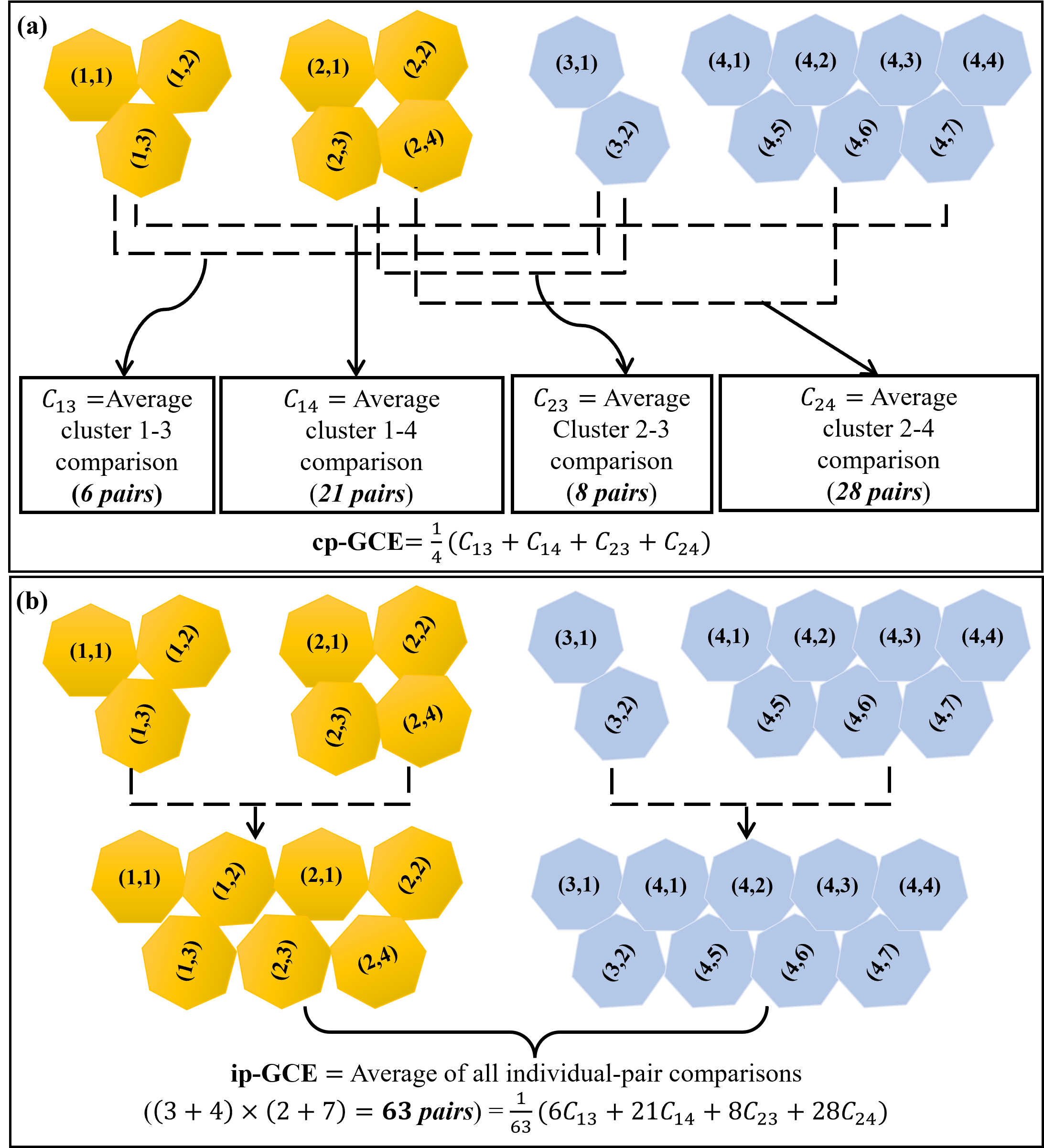}
    \caption{Schematic of the (a) cp-GCE and (b) ip-GCE definitions. In each panel, $(i,j)$ is individual $j$ in cluster $i$; the two yellow (blue) clusters are independent random draws from the population of clusters with $A_i=1$ ($A_i=0$).}
    \vspace*{-0.2in}
    \label{fig:win_scheme}
\end{figure}

Three features are worth highlighting. {\color{black}\label{par:cpip}First, the two estimands differ only in their weighting scheme. The cp-GCE in \eqref{eq:estimand-C} weights every cluster pair equally regardless of size and so targets the population of clusters, whereas the ip-GCE in \eqref{eq:estimand-I} weights every individual pair equally and so targets the population of individuals irrespective of cluster membership; this generalizes the distinction of \citet{kahan2023estimands} from singletons to pairs and is what accommodates informative cluster sizes. The choice should follow the scientific question and intended unit of inference. The cp-GCE $\lambda_{C,a}$ is relevant when interest centers on comparisons between equally weighted clusters, such as evaluating provider performance with small and large clinics given equal importance. This target is particularly relevant to CRTs with a cluster-level intervention and less natural in individually randomized trials without an explicit cluster structure. The ip-GCE $\lambda_{I,a}$ instead concerns comparisons between individuals drawn from the population, resembling the target of an individually randomized trial, and may be preferred when cluster randomization of an individual-level intervention is adopted primarily for logistical reasons \citep{li2025model}. These are guiding examples rather than strict rules, and both estimands may be scientifically informative.} Second, both are defined through cross-cluster rather than within-cluster comparisons, because within-cluster contrasts such as $\bbE[\sum_{j=1}^{N_1} w\{\bfY_{1j}(a),\bfY_{1j}(1-a)\}/N_1]$ are not estimable for nonlinear $w$, by the fundamental problem of causal inference \citep{Mao2018}. Third, when cluster sizes vary only at random, the two coincide such that $\lambda_{C,a}=\lambda_{I,a}=\bbE[w\{\bfY_{1j}(a),\bfY_{2l}(1-a)\}]$. Figure \ref{fig:win_scheme} illustrates both definitions schematically.

\begin{remark}\label{rmk:estimand}
    Our GCE estimands serve as building blocks for a broader class of summary causal estimands of the form $\Lambda = f(\lambda_1,\lambda_0)$, where $f(\cdot,\cdot)$ is a known function defining the scale of measure. Throughout, $f$ is continuously differentiable at the true $(\lambda_1,\lambda_0)$, and the asymptotic variances of the estimators of $\Lambda$ below are positive. Suppose $w(\bfu,\bfv) = \bbone(\bfu \succ \bfv)$ for a comparison rule as in Example \ref{ex:prioritized}. If we specify $f(u,v)=u-v$, then $\Lambda_C=\lambda_{C,1}-\lambda_{C,0}$ and $\Lambda_I=\lambda_{I,1}-\lambda_{I,0}$ become the cluster-pair and individual-pair net benefit estimands. If we specify $f(u,v)=u/v$, $\Lambda_C=\lambda_{C,1}/\lambda_{C,0}$ and $\Lambda_I=\lambda_{I,1}/\lambda_{I,0}$ become the cluster-pair and individual-pair win ratio estimands. Furthermore, if we modify the contrast function to give half credit to ties, $w(\bfu,\bfv) = \bbone(\bfu \succ \bfv)+0.5\{1-\bbone(\bfu \succ \bfv)-\bbone(\bfv \succ \bfu)\}$, and set $f(u,v)=u/v$, then $\Lambda_C=\lambda_{C,1}/\lambda_{C,0}$ and $\Lambda_I=\lambda_{I,1}/\lambda_{I,0}$ become the cluster-pair and individual-pair win odds estimands. These estimands extend their counterparts defined for independent data \citep{buyse2010generalized,bebu2016large,brunner2021win} to the CRT setting.
\end{remark}

\begin{remark}
    When $Q=1$ and a single outcome is considered, our estimands nest two special cases. First, if we specify the contrast function as a difference function $w(Y_{ij},Y_{kl})=Y_{ij}-Y_{kl}$, then $\lambda_{C,a}$ and $\lambda_{I,a}$ reduce to the conventional cluster-average and individual-average treatment effect estimands \citep{kahan2023estimands}. Second, if we specify the contrast as $w(Y_{ij},Y_{kl})=\bbone(Y_{ij}>Y_{kl})+0.5\bbone(Y_{ij}=Y_{kl})$, then $\lambda_{C,a}$ and $\lambda_{I,a}$ become the cluster-average and individual-average probabilistic index estimands, generalizing the counterpart used previously for independent data \citep{thas2012probabilistic}.
\end{remark}

For cluster $i$, we define the full data as $\calD_i=\{\mathbf Y_i(1),\mathbf Y_i(0),N_i,\bfC_i,\mathbf X_i\}$, and the observed data as $\calO_i=\{\mathbf Y_i,A_i,N_i,\bfC_i,\mathbf X_i\}$, where $\mathbf Y_i=\{\bfY_{ij},j=1,\ldots,N_i\}$ and $\mathbf X_i=\{\bfX_{ij},j=1,\ldots,N_i\}$. We next introduce two additional structural assumptions.

\begin{assumption}[Population distributions] \label{asp:super-pop}
    (i) $\calD_1,\ldots,\calD_m$ are mutually independent. (ii) The unknown cluster size distribution $\calP(N)$ is defined over a finite support on $\bbN^+$. (iii) Given $N_i$, $\calD_i$ follows an unknown distribution $\calP(\calD|N)$ with finite second moments.
\end{assumption}

\begin{assumption}[Cluster-level randomization] \label{asp:cl-ran}
    For cluster $i$, the treatment indicator $A_i$ is an independent draw from the Bernoulli distribution $\calP(A)$ with known parameter $\bbP(A_i = 1) = \pi\in(0, 1)$. Furthermore, $\{A_1,\ldots,A_m\}$ are independent of $\{\calD_1,\ldots,\calD_m\}$.
\end{assumption}

Assumption \ref{asp:super-pop}(i) requires that the variables across clusters are independent, but outcomes and covariates within the same cluster can be arbitrarily correlated. Assumption \ref{asp:super-pop}(ii)-(iii) state that $\calD_1, \ldots,\calD_m$ are marginally identically distributed according to a mixture distribution, $\calP(\calD|N)\calP(N)$. Hence, Assumption \ref{asp:super-pop} depicts the super-population framework and underlies our asymptotic regimes introduced in Section \ref{sec:est}. Assumption \ref{asp:cl-ran} describes Bernoulli cluster randomization and holds by design. 

\begin{remark} \label{rmk:within-cluster-subsampling}
    We primarily consider the case in which the observed members in each cluster constitute the cluster's population of interest. In other cases, however, we may observe data from only $0<M_i\leq N_i$ units in cluster $i$, and the observed cluster size may depend on assignment and cluster-level attributes \citep{wang2024model}. That is, $M_i=A_iM_i(1)+(1-A_i)M_i(0)$ with $M_i(a)=h_a(N_i,\bfC_i,\epsilon_{a,i})$, where $h_a$ is a treatment-dependent unknown function and $\epsilon_{a,i}$ is a treatment-dependent noise independent from $\{N_i,\bfC_i,\mathbf X_i,\mathbf Y_i(a)\}$. We refer to this setting as within-cluster subsampling and provide an extension of our main theoretical results under this setting in Section S9 of the Supplementary Materials. 
\end{remark}

\section{Estimating the generalized causal effects} \label{sec:est}

\subsection{Nonparametric moment estimators} \label{subsec:np-estimator}

Equations \eqref{eq:estimand-C}--\eqref{eq:estimand-I} lead to the moment conditions
\begin{align*}
    &\bbE\left[\frac{\sum_{j=1}^{N_1}\sum_{l=1}^{N_2} w\{\bfY_{1j}(a),\bfY_{2l}(1-a)\}}{N_1N_2}-\lambda_{C,a}\right]=0\quad\text{and}\\
    &\bbE\left[\sum_{j=1}^{N_1}\sum_{l=1}^{N_2} w\{\bfY_{1j}(a),\bfY_{2l}(1-a)\}-N_1N_2\lambda_{I,a}\right]=0,
\end{align*}
which motivate the following U-estimating equations
\begin{equation} \label{eq:np-equations}
    \sum_{1\leq i<k\leq m}\bfpsi_v^\np(\calO_i,\calO_k;\bflambda) = \sum_{1\leq i<k\leq m}\left\{\begin{array}{l}
        \psi_{v,1}^\np(\calO_i,\calO_k;\lambda_1) \\[1ex]
        \psi_{v,0}^\np(\calO_i,\calO_k;\lambda_0)
    \end{array}\right\} = \bm 0,
\end{equation}
where $v=C,I$. For $a\in\{0,1\}$,
\begin{align*}
    \psi_{C,a}^\np(\calO_i,\calO_k;\lambda_a)&=2^{-1}\left[\frac{\bbone(A_i=a)\bbone(A_k=1-a)}{N_iN_k}\sum_{j=1}^{N_i}\sum_{l=1}^{N_k}\left\{w(\bfY_{ij},\bfY_{kl})-\lambda_a\right\} \right.\displaybreak[0]\\
    &\qquad\qquad\left.+ \frac{\bbone(A_k=a)\bbone(A_i=1-a)}{N_iN_k}\sum_{j=1}^{N_i}\sum_{l=1}^{N_k}\left\{w(\bfY_{kl},\bfY_{ij})-\lambda_a\right\}\right]
\end{align*}
and $\bfpsi_I^\np(\calO_i,\calO_k;\bflambda)=N_iN_k\bfpsi_C^\np(\calO_i,\calO_k;\bflambda)$. Solving \eqref{eq:np-equations} then gives the nonparametric estimator $\widehat\lambda_{v,a}^\np$, and the nonparametric estimator for $\Lambda_v$ is $\widehat\Lambda_v^\np=f(\widehat\lambda_{v,1}^\np,\widehat\lambda_{v,0}^\np)$. 

For inference, let $\bfB_v^\np=\bbE\{\nabla_{\bflambda}\bfpsi_v^\np(\calO_i,\calO_k;\bflambda_v)\}$ denote the Jacobian, where $\nabla_{\bflambda}\bfpsi_v^\np(\calO_i,\calO_k;\bflambda_v)$ is the first-order derivative of $\bfpsi_v^\np(\calO_i,\calO_k;\bflambda)$ evaluated at $\bflambda_v=(\lambda_{v,1},\lambda_{v,0})^\top$. The asymptotic covariance matrix of $m^{1/2}\widehat\bflambda_v^\np=m^{1/2}(\widehat\lambda_{v,1}^\np,\widehat\lambda_{v,0}^\np)^\top$ is $\bfV_v^\np=(\bfB_v^\np)^{-1}\bfSigma_v^\np\{(\bfB_v^\np)^{-1}\}^\top$, where $\bfSigma_v^\np=4\Var\{\overline\bfpsi_v^\np(\calO_i;\allowbreak\bflambda_v)\}$ and $\overline\bfpsi_v^\np(\calO_i;\bflambda)=\bbE\{\bfpsi_v^\np(\calO_i,\calO_k;\bflambda)|\calO_i\}$ is the H\'ajek projection of the estimating function \citep[\S11]{vandervaart1998}. The asymptotic covariance matrix $\bfV_v^\np$ is estimated by the sandwich variance estimator $\widehat\bfV_v^\np=(\widehat \bfB_v^\np)^{-1}\widehat\bfSigma_v^\np\{(\widehat \bfB_v^\np)^{-1}\}^\top$. {\color{black}For brevity, the explicit forms for the empirical Jacobian $\widehat \bfB_v^\np$ and the sample variance of the estimated H\'ajek projections $\widehat\bfSigma_v^\np$ are provided in Section S3 of the Supplementary Materials.} Next, for the summary estimand in Remark \ref{rmk:estimand}, we let $V_v^\np$ denote the asymptotic variance of $m^{1/2}\widehat\Lambda_v^\np$, with $\widehat V_v^\np$ denoting its estimator. Then, by the Delta method, $V_v^\np=\{\nabla_{\bflambda} f(\bflambda_v)\}^\top\bfV_v^\np\nabla_{\bflambda} f(\bflambda_v)$, which is estimated by $\widehat V_v^\np=\{\nabla_{\bflambda} f(\widehat\bflambda_v^\np)\}^\top\widehat\bfV_v^\np\nabla_{\bflambda} f(\widehat\bflambda_v^\np)$. We first give the following theorem, which summarizes the asymptotic properties of the nonparametric estimators.
\begin{theorem} \label{thm:np-est}
    Under Assumptions \ref{asp:sutva}--\ref{asp:cl-ran}, for $v=C,I$, if regularity conditions (P1)--(P4) in Section S2 of the Supplementary Materials hold, then $\widehat\bflambda_v^\np\overset{p}{\to}\bflambda_v$; furthermore, if regularity conditions (P5)--(P7) also hold, then $m^{1/2}(\widehat\bflambda_v^\np-\bflambda_v)\overset{d}{\to}\calN(\bm 0,\bfV_v^\np)$ and $\widehat \bfV_v^\np\overset{p}{\to}\bfV_v^\np$.
\end{theorem}

Theorem \ref{thm:np-est} implies that $m^{1/2}(\widehat V_v^\np)^{-1/2}(\widehat\Lambda_v^\np-\Lambda_v)\overset{d}{\to}\calN(0,1)$ for $v=C,I$, and follows from the law of large numbers and the central limit theorem for U-statistics \citep[\S12]{vandervaart1998}. Conditions (P1)--(P7) are those routinely adopted in the M- and Z-estimation literature \citep[\S5]{vandervaart1998}. Specifically, (P1)--(P4) are for consistency and require identifiability of $\bflambda_v$, Lipschitz continuity of $\bfpsi_v^\np$ in $\bflambda$ with a square-integrable envelope, and finiteness of $\bbE\{\|\bfpsi_v^\np\|^2\}$; (P5)--(P7) are for asymptotic normality and consistency of $\widehat\bfV_v^\np$, and require twice-differentiability of $\bfpsi_v^\np$ in $\bflambda$, existence of $\bfB_v^\np$ with a square-integrable derivative, and Lipschitz continuity of $\nabla_{\bflambda}\bfpsi_v^\np$ in $\bflambda$ with a square-integrable envelope.

\subsection{The efficient influence functions} \label{subsec:eif}

To improve upon nonparametric moment estimators, we first derive EIFs for $\lambda_{C,a}$ and $\lambda_{I,a}$, which motivate estimators that leverage baseline covariates to enhance efficiency without compromising consistency. By the semiparametric efficiency theory \citep[\S4]{Tsiatis2006}, the estimators constructed based on EIFs can attain the semiparametric efficiency lower bound for $\lambda_{C,a}$ and $\lambda_{I,a}$, under appropriate conditions, and are optimal. {\color{black} Let $\mathbf W_i = \{\mathbf X_i, \bfC_i, N_i\}$ denote the complete baseline covariate collection for cluster $i$.} Defining $\overline w(\mathbf Y_i,\mathbf Y_k)=(N_iN_k)^{-1}\sum_{j=1}^{N_i}\sum_{l=1}^{N_k}w(\bfY_{ij},\bfY_{kl})$ as the observed average contrast between two clusters regardless of treatment assignment, the EIFs for $\lambda_{C,a}$ and $\lambda_{I,a}$ are given in Theorem \ref{thm:eif}.

\begin{theorem} \label{thm:eif}
    Suppose Assumptions \ref{asp:sutva}--\ref{asp:cl-ran} hold and $\bbE[\overline w\{\mathbf Y_1(a),\mathbf Y_2(1-a)\}^2]<\infty$ for $a\in\{0,1\}$.
    \begin{itemize}
        \item[(i)] The EIF for $\lambda_{C,a}$ is
        \begin{align*}
            &\varphi_{C,a}(\calO)=\frac{\bbone(A=a)}{\pi^a(1-\pi)^{1-a}}\left[\frakw_{C,a,a}(\mathbf Y)-h_{C,a,a}(\mathbf W)\right]\displaybreak[0]\\
            &~+\frac{\bbone(A=1-a)}{\pi^{1-a}(1-\pi)^a}\left[\frakw_{C,a,1-a}(\mathbf Y)-h_{C,a,1-a}(\mathbf W)\right]+h_{C,a,a}(\mathbf W)+h_{C,a,1-a}(\mathbf W)-2\lambda_{C,a},
        \end{align*}
        where $\frakw_{C,a,a}(\mathbf y)=\bbE\{\overline w(\mathbf y,\mathbf Y_k)|A_k=1-a\}$, $\frakw_{C,a,1-a}(\mathbf y)=\bbE\{\overline w(\mathbf Y_k,\mathbf y)|A_k=a\}$ and $h_{C,a,a}(\mathbf W)=\bbE\{\frakw_{C,a,a}(\mathbf Y)|\mathbf W,A=a\}$, $h_{C,a,1-a}(\mathbf W)=\bbE\{\frakw_{C,a,1-a}(\mathbf Y)|\mathbf W,A=1-a\}$.
        \item[(ii)] The EIF for $\lambda_{I,a}$ is
        \begin{align*}
            &\varphi_{I,a}(\calO)=\frac{\bbone(A=a)N}{\pi^a(1-\pi)^{1-a}\{\bbE(N)\}^2}\left[\frakw_{I,a,a}(\mathbf Y)-h_{I,a,a}(\mathbf W)\right]\displaybreak[0]\\
            &~+\frac{\bbone(A=1-a)N}{\pi^{1-a}(1-\pi)^a\{\bbE(N)\}^2}\left[\frakw_{I,a,1-a}(\mathbf Y)-h_{I,a,1-a}(\mathbf W)\right]\displaybreak[0]\\
            &~+\frac{N}{\{\bbE(N)\}^2}\left[h_{I,a,a}(\mathbf W)+h_{I,a,1-a}(\mathbf W)\right] -\frac{2N\lambda_{I,a}}{\bbE(N)},
        \end{align*}
        where $\frakw_{I,a,a}(\mathbf y)=\bbE\{N_k\overline w(\mathbf y,\mathbf Y_k)|A_k=1-a\}$, $\frakw_{I,a,1-a}(\mathbf y)=\bbE\{N_k\overline w(\mathbf Y_k,\mathbf y)|A_k=a\}$ and $h_{I,a,a}(\mathbf W)=\bbE\{\frakw_{I,a,a}(\mathbf Y)|\mathbf W,A=a\}$, $h_{I,a,1-a}(\mathbf W)=\bbE\{\frakw_{I,a,1-a}(\mathbf Y)|\mathbf W,A=1-a\}$.
    \end{itemize}
\end{theorem}

The EIFs in Theorem \ref{thm:eif} give the contribution of each cluster to the estimation. This leads to the functions $\frakw_{v,a,a}(\mathbf y)$ and $\frakw_{v,a,1-a}(\mathbf y)$ for $v=C,I$, which are essentially half-marginalized observed contrasts, and are to be estimated using empirical averages. Consequently, the specific structures of the EIFs directly motivate the construction of the subsequent U-estimating equations in Section \ref{subsec:mr-estimator}, whose H\'ajek projections onto singleton observations recover the EIFs. With independent data ($N=1$), the two EIFs coincide and reduce to the optimal estimating function in \citet{Mao2018} with known treatment assignment. More generally, $\varphi_{C,a}(\calO)$ and $\varphi_{I,a}(\calO)$ accommodate correlated data with varying cluster sizes. The semiparametric efficiency bounds for $\lambda_{C,a}$ and $\lambda_{I,a}$ are $\bbE\{\varphi_{C,a}(\calO)^2\}$ and $\bbE\{\varphi_{I,a}(\calO)^2\}$, respectively. The EIFs and efficiency bounds for any derived estimands from $\lambda_{C,a}$ and $\lambda_{I,a}$ can then be obtained from Theorem \ref{thm:eif} via the chain rule. {\color{black}Finally, the EIFs also characterize the efficiency gain from baseline covariate adjustment. Following \citet{Kallus2025}, we compare the asymptotic variance $V_{v,a}^{\np}$ of $m^{1/2}\widehat\lambda_{v,a}^{\np}$ with the semiparametric efficiency bound $V_{v,a}^{\eff}=\bbE\{\varphi_{v,a}(\calO)^2\}$. Write $\pi_a=\pi^a(1-\pi)^{1-a}$. Under the conditions of Theorem \ref{thm:eif}, for $a\in\{0,1\}$, we prove in Section S4.1 of the Supplementary Materials that
\begin{align}
    V_{C,a}^{\np} - V_{C,a}^{\eff} &= \pi(1-\pi)\,\bbE\left[\left\{\frac{h_{C,a,a}(\mathbf W)-\lambda_{C,a}}{\pi_a} - \frac{h_{C,a,1-a}(\mathbf W)-\lambda_{C,a}}{\pi_{1-a}}\right\}^2\right]\geq 0, \label{eq:effgain-C}\displaybreak[0]\\
    V_{I,a}^{\np} - V_{I,a}^{\eff} &= \frac{\pi(1-\pi)}{\{\bbE(N)\}^4}\,\bbE\left[N^2\left\{\frac{h_{I,a,a}(\mathbf W)-\bbE(N)\lambda_{I,a}}{\pi_a} - \frac{h_{I,a,1-a}(\mathbf W)-\bbE(N)\lambda_{I,a}}{\pi_{1-a}}\right\}^2\right]\geq 0. \label{eq:effgain-I}
\end{align}
The efficiency gain is the second moment of a weighted contrast between the centered conditional mean functions, which capture the components of the half-marginalized pairwise contrasts predictable from $\mathbf W$. For the cluster-pair estimand, the gain is zero if and only if $h_{C,a,a}(\mathbf W)/\pi_a-h_{C,a,1-a}(\mathbf W)/\pi_{1-a}$ is constant almost surely, as occurs when covariates carry no predictive information about the cross-cluster contrasts, and otherwise increases with its variance. For the individual-pair estimand, the additional factor $N^2/\{\bbE(N)\}^4$ weights the corresponding squared contrast more heavily in larger clusters, so the efficiency gain depends jointly on covariate predictiveness and the cluster size distribution.}

\subsection{Model-robust estimators} \label{subsec:mr-estimator}

The EIFs in Theorem \ref{thm:eif} suggest approaches to leverage baseline covariates for efficiency improvement. Covariates are incorporated into these estimators through the nuisance function, i.e., the outcome mean. Theorem \ref{thm:mr} below shows that these EIF-motivated estimators are robust to misspecification of the working nuisance model. {\color{black} To begin, we define $\zeta_{ik,a}=\zeta_a(\mathbf W_i,\mathbf W_k)=\bbE\{\overline w(\mathbf Y_i,\mathbf Y_k)|A_i=a,A_k=1-a,\mathbf W_i,\mathbf W_k\}$}, the expected average contrast between clusters $i$ and $k$ with cluster $i$ under treatment $a$ and cluster $k$ under $1-a$, conditional on observed individual- and cluster-level covariates as well as cluster sizes.

We next consider a working model for $\zeta_a$. That is, $\zeta_a = \zeta_a(\bfvartheta_a)$, where $\zeta_a$ is a prespecified function with corresponding finite-dimensional parameter $\bfvartheta_a$. Define the estimated nuisance function $\widehat\zeta_a$ and the true nuisance function $\zeta_a^0$, and, in this case, $\widehat\zeta_a=\zeta_a(\widehat\bfvartheta_a)$. The EIFs in Theorem \ref{thm:eif} motivate the following U-estimating equations:
\begin{equation} \label{eq:dr-equations}
    \sum_{1\leq i<k\leq m}\bfpsi_v^\eff(\calO_i,\calO_k;\widehat\bfvartheta,\bflambda) = \sum_{1\leq i<k\leq m}\left\{\begin{array}{l}
        \psi_{v,1}^\eff(\calO_i,\calO_k;\widehat\bfvartheta,\lambda_1) \\[1ex]
        \psi_{v,0}^\eff(\calO_i,\calO_k;\widehat\bfvartheta,\lambda_0)
    \end{array}\right\} = \bm 0,
\end{equation}
where $v=C,I$. For $a\in\{0,1\}$,
\begin{align*}
    &\psi_{C,a}^\eff(\calO_i,\calO_k;\widehat\bfvartheta,\lambda_a)=2^{-1}\Bigg[\frac{\bbone(A_i=a)\bbone(A_k=1-a)}{\pi(1-\pi)}\left\{\overline w(\mathbf Y_i,\mathbf Y_k)- \zeta_{ik,a}(\widehat\bfvartheta_a)\right\}\displaybreak[0]\\
    &+\frac{\bbone(A_k=a)\bbone(A_i=1-a)}{\pi(1-\pi)}\left\{\overline w(\mathbf Y_k,\mathbf Y_i)- \zeta_{ki,a}(\widehat\bfvartheta_a)\right\}+\zeta_{ik,a}(\widehat\bfvartheta_a)+\zeta_{ki,a}(\widehat\bfvartheta_a)-2\lambda_a\Bigg]
\end{align*}
and $\bfpsi_I^\eff(\calO_i,\calO_k;\widehat\bfvartheta,\bflambda)=N_iN_k\bfpsi_C^\eff(\calO_i,\calO_k;\widehat\bfvartheta,\bflambda)$. Solving \eqref{eq:dr-equations} directly leads to estimator $\widehat\bflambda_v^\mr=(\widehat\lambda_{v,1}^\mr,\widehat\lambda_{v,0}^\mr)^\top$. Let $\bfV_v^\mr$ denote the asymptotic covariance matrix of $m^{1/2}\widehat\bflambda_v^\mr$ with sandwich variance estimator $\widehat \bfV_v^\mr$, where explicit expressions of $\bfV_v^\mr$ and $\widehat \bfV_v^\mr$ are given in Section S5.2 of the Supplementary Materials. Also, let $\underline\bfvartheta_a$ denote the probability limit of $\widehat\bfvartheta_a$. The following Theorem summarizes the properties of the model-robust estimators.
\begin{theorem} \label{thm:mr}
    Under Assumptions \ref{asp:sutva}--\ref{asp:cl-ran}, for $v=C,I$, if regularity conditions (P1)-(P4) in Section S2 of the Supplementary Materials hold, then $\widehat\bflambda_v^\mr\overset{p}{\to}\bflambda_v$; if regularity conditions (P5)--(P7) also hold, then $m^{1/2}(\widehat\bflambda_v^\mr-\bflambda_v)\overset{d}{\to}\calN(\bm 0,\bfV_v^\mr)$ and $\widehat \bfV_v^\mr\overset{p}{\to}\bfV_v^\mr$; these results hold regardless of the correct specification of the nuisance model $\zeta_a$. Moreover, if $\zeta_a(\underline\bfvartheta_a)=\zeta_a^0$, then $\widehat \bfV_v^\mr$ converges in probability to the semiparametric efficiency lower bound of $\bflambda_v$.
\end{theorem}

Since $\zeta_a = \zeta_a(\bfvartheta_a)$ is a working model, the parameter $\bfvartheta_a$ can be estimated by solving U-estimating equations $\sum_{1\leq i<k\leq m}\bfpsi_{\bfvartheta}(\calO_i,\calO_k;\bfvartheta)=\bm 0$, where $\bfpsi_{\bfvartheta}=(\bfpsi_{\bfvartheta,1},\bfpsi_{\bfvartheta,0})$. Hence, a joint estimation equation for $(\bfvartheta,\bflambda_v)$ can be formed as $\sum_{1\leq i<k\leq m}(\bfpsi_{\bfvartheta}, \bfpsi_v^\eff)=\bm 0$, where $\bfpsi_v^\eff$ is given in \eqref{eq:dr-equations}, leading to the applicability of the same set of regularity conditions (P1)--(P7). The estimator $\widehat\lambda_{v,a}^\mr$ is model-robust in the sense that it is consistent and asymptotically normal regardless of whether $\zeta_a(\underline\bfvartheta_a)$ correctly specifies $\zeta_a^0$. When $\zeta_a(\underline\bfvartheta_a)=\zeta_a^0$, twice the H\'ajek projection of $\psi_{v,a}^\eff$, divided by $\{\bbE(N)\}^2$ if $v=I$, is the EIF, and $\bfV_v^\mr$ attains the semiparametric efficiency bound of $\bflambda_v$. For $\widehat\Lambda_v^\mr=f(\widehat\lambda_{v,1}^\mr,\widehat\lambda_{v,0}^\mr)$ and $\widehat V_v^\mr=\{\nabla_{\bflambda} f(\widehat\bflambda_v^\mr)\}^\top\widehat\bfV_v^\mr\nabla_{\bflambda} f(\widehat\bflambda_v^\mr)$, Theorem \ref{thm:mr} implies that, for $v=C,I$, $m^{1/2}(\widehat V_v^\mr)^{-1/2}(\widehat\Lambda_v^\mr-\Lambda_v)\overset{d}{\to}\calN(0,1)$, and, if $\zeta_a(\underline\bfvartheta_a)=\zeta_a^0$, then $\widehat V_v^\mr$ converges in probability to the semiparametric efficiency lower bound of $\Lambda_v$.  

The model-robust estimator $\widehat\lambda_{v,a}^\mr$ is a U-statistic analog of the augmented inverse probability weighted (AIPW) estimator for conventional average treatment effect estimands in CRTs \citep{li2025model}; {\color{black}the two are placed side by side in Section S5.1 of the Supplementary Materials.} One implementation caveat deserves emphasis. That is, unlike the AIPW estimator for the average treatment effects, which requires two conditional expectations, $\widehat\lambda_{v,1}^\mr$ and $\widehat\lambda_{v,0}^\mr$ require ``imputing'' all four conditional expectations listed in Table \ref{tab:conditional-expectations}, because within a pair either member may occupy the winning position.

\begin{table}[htbp]
    \centering
    \caption{The four conditional expectations to be imputed when implementing the model-robust estimator for the cp/ip-GCE estimands, contrasted with the two required by the AIPW estimator for the c/i-ATE. The combination of treatment $a$ and observation $i$ means that observation $i$ is in the winning position and under treatment $a$; $\bfU_i$ denotes user-specified covariate vectors that are arbitrary functions of $(A_i,\mathbf W_i)$.} \label{tab:conditional-expectations}
    \begin{tabular}{cccc}
        \toprule
        & & \multicolumn{2}{c}{Estimands} \\
        \cmidrule{3-4}
        Treatment & Obs. & c/i-ATE & cp/ip-GCE \\
        \midrule
        \multirow{2}{*}{$a$} & $i$ & $\bbE(\overline Y_i|A_i=a,\bfU_i)$ & $\bbE\{\overline w(\mathbf Y_i,\mathbf Y_k)|A_i=a,A_k=1-a,\bfU_i,\bfU_k\}$ \\
        & $k$ & $-$ & $\bbE\{\overline w(\mathbf Y_k,\mathbf Y_i)|A_k=a,A_i=1-a,\bfU_k,\bfU_i\}$ \\
        \multirow{2}{*}{$1-a$} & $i$ & $\bbE(\overline Y_i|A_i=1-a,\bfU_i)$ & $\bbE\{\overline w(\mathbf Y_i,\mathbf Y_k)|A_i=1-a,A_k=a,\bfU_i,\bfU_k\}$ \\
        & $k$ & $-$ & $\bbE\{\overline w(\mathbf Y_k,\mathbf Y_i)|A_k=1-a,A_i=a,\bfU_k,\bfU_i\}$  \\
        \bottomrule
    \end{tabular}
\end{table}

Because $\mathbf X_i$ and $\mathbf X_k$ collect covariate vectors of varying dimension, a practical strategy is to model $w(\bfY_{ij},\bfY_{kl})$ on $(\bfX_{ij},\bfX_{kl},\bfC_i,\bfC_k,N_i,N_k)$ together with fixed-dimensional summaries of $\mathbf X_i$ and $\mathbf X_k$, such as the cluster averages $\overline{\mathbf X}_i$ and $\overline{\mathbf X}_k$, and then average the predictions within the cluster pair. Let $\bfU_{ij}$ denote a user-specified covariate vector that is a pre-specified function of $(A_i,\mathbf W_i,\bfX_{ij})$, {\color{black}with dimension $p_u$}. The four conditional expectations in Table \ref{tab:conditional-expectations} can then be imputed through a pairwise specification such as the probabilistic index model (PIM) \citep{thas2012probabilistic}, a generalized linear model $\bbE\{w(\bfY_{ij},\bfY_{kl})|\bfU_{ij},\bfU_{kl}\}=g^{-1}\{(\bfU_{ij}-\bfU_{kl})^\top\bfvartheta\}$ with a known link $g$, fitted to all pairs $(i,j)$ and $(k,l)$. Averaging its fitted values over $j$ and $l$ gives $\zeta_{ik,a}$. In $\zeta_{ik,a}$, the treatment components of $\bfU_{ij}$ and $\bfU_{kl}$ are set to $a$ and $1-a$. We give a concrete example below.

\begin{example}\label{ex:PIM}
    Let $w(\bfY_{ij},\bfY_{kl})=\bbone(\bfY_{ij}\preceq_P \bfY_{kl})$ be the Pareto partial order of Example \ref{ex:pareto} and $\bfU_{ij}=(A_i,\bfX_{ij}^\top,\overline{\mathbf X}_i^\top,\bfC_i^\top, N_i)^\top$. The PIM working model for $\zeta_{ik,a}$ can be specified as
    \begin{align*}
        g[\bbE\{\bbone(\bfY_{ij}\preceq_P \bfY_{kl})|\bfU_{ij},\bfU_{kl}\}] &= (A_i-A_k)\beta_A + (\bfX_{ij}-\bfX_{kl})^\top\bfbeta_{\bfX}\displaybreak[0]\\
        &\quad+(\overline{\mathbf X}_i-\overline{\mathbf X}_k)^\top\bfbeta_{\overline{\mathbf X}} + (\bfC_i-\bfC_k)^\top\bfbeta_{\bfC} + (N_i-N_k)\beta_N,
    \end{align*}
    with $A_i=a$, $A_k=1-a$, and $\bfvartheta = (\beta_A, \bfbeta_{\bfX}^\top, \bfbeta_{\overline{\mathbf X}}^\top, \bfbeta_{\bfC}^\top, \beta_N)^\top$. Fitting this model via the routine of \citet{thas2012probabilistic} gives $\widehat\bfvartheta$, and the nuisance estimate is obtained by averaging the fitted values within the pair, $\widehat\zeta_{ik,a}=(N_iN_k)^{-1}\sum_{j=1}^{N_i}\sum_{l=1}^{N_k}g^{-1}\{(\bfU_{ij}-\bfU_{kl})^\top\widehat\bfvartheta\}$.
\end{example}

When $Q=1$ and $w(u,v)=\bbone(u>v)$, a simpler listwise specification is available. Under a working Gaussian mixed model $Y_{ij}(a)\sim\calN(f_a(\bfU_{ij}),\sigma^2)$ with $f_a(\bfU_{ij})=\bfU_{ij}^\top\bfvartheta_a$, the induced pairwise mean is $\zeta_a(\bfU_{ij},\bfU_{kl};\bfvartheta)=\Phi[\{f_a(\bfU_{ij})-f_{1-a}(\bfU_{kl})\}/(\sqrt{2}\sigma)]$ \citep{wolfe1971constructing}.

\subsection{The debiased machine learning estimators} \label{subsec:dml-estimator}

With the EIFs, we further leverage data-adaptive machine learning methods with cross-fitting to estimate nuisance functions, yielding asymptotically efficient DML estimators \citep{Chernozhukov2018}. For cross-fitting, we extend the scheme in \citet{escanciano2023machine} and \citet{chen2024principal} to clustered U-statistics. Specifically, the single index set $\{1,\ldots,m\}$ is first divided into $|\frakI^\si|$ subsets of similar sizes, forming a partition $\frakI^\si=\{\calI_1^\si,\ldots,\calI_{|\frakI^\si|}^\si\}$. Then, upon $\frakI^\si$, a second partition, $\frakI^\pr$, can be formed for all pairs $(i,k)$ with $1\leq i<k\leq m$, which has $|\frakI^\pr|=|\frakI^\si|(|\frakI^\si|+1)/2$ elements, i.e., $\frakI^\pr=\{\calI_1^\pr,\ldots,\calI_{|\frakI^\pr|}^\pr\}$. \citet{Chernozhukov2018} suggested setting $|\frakI^\si|$ to moderate values, such as four or five, so that $|\frakI^\pr|$ does not grow with $m$. An example demonstrating the structures of $\frakI^\si$ and $\frakI^\pr$ is given in Section S6 of the Supplementary Materials.

We obtain the following U-estimating equations for the DML estimators:
\begin{equation} \label{eq:dml-equations}
    \sum_{p=1}^{|\frakI^\pr|}\sum_{(i,k)\in \calI_p^\pr}\bfpsi_v^\eff(\calO_i,\calO_k;\widehat \zeta_p,\bflambda) = \sum_{p=1}^{|\frakI^\pr|}\sum_{(i,k)\in \calI_p^\pr}\left\{\begin{array}{l}
        \psi_{v,1}^\eff(\calO_i,\calO_k;\widehat \zeta_p,\lambda_1) \\[1ex]
        \psi_{v,0}^\eff(\calO_i,\calO_k;\widehat \zeta_p,\lambda_0)
    \end{array}\right\} = \bm 0,
\end{equation}
for $v=C,I$. For $a\in\{0,1\}$,
\begin{align*}
    &\psi_{C,a}^\eff(\calO_i,\calO_k;\widehat\zeta_p,\lambda_a)=2^{-1}\Bigg[\frac{\bbone(A_i=a)\bbone(A_k=1-a)}{\pi(1-\pi)}\left\{\overline w(\mathbf Y_i,\mathbf Y_k)- \widehat\zeta_{ik,p,a}\right\}\displaybreak[0]\\
    &\qquad+\frac{\bbone(A_k=a)\bbone(A_i=1-a)}{\pi(1-\pi)}\left\{\overline w(\mathbf Y_k,\mathbf Y_i)- \widehat\zeta_{ki,p,a}\right\}+\widehat\zeta_{ik,p,a}+\widehat\zeta_{ki,p,a}-2\lambda_a\Bigg]
\end{align*}
and $\bfpsi_I^\eff(\calO_i,\calO_k;\widehat\zeta_p,\bflambda)=N_iN_k\bfpsi_C^\eff(\calO_i,\calO_k;\widehat\zeta_p,\bflambda)$. For a given partition $\frakI^\pr$, we form intermediate estimating equations using observations in each $\calI_p^\pr$ with $\widehat\zeta_{p,a}$ estimated using observations in $(\calI_p^\pr)^c$, comprised of clusters not in $\calI_p^\pr$. Solving \eqref{eq:dml-equations} leads to the DML estimators $\widehat\bflambda_v^\dml=(\widehat\lambda_{v,1}^\dml,\widehat\lambda_{v,0}^\dml)^\top$ and $\widehat\Lambda_v^\dml=f(\widehat\lambda_{v,1}^\dml,\widehat\lambda_{v,0}^\dml)$. {\color{black} Let $\widehat\bfV_v^\dml=(\widehat\bfB_v^\dml)^{-1}\widehat\bfSigma_v^\dml\{(\widehat\bfB_v^\dml)^{-1}\}^\top$ and $\widehat V_v^\dml=\{\nabla_{\bflambda} f(\widehat\bflambda_v^\dml)\}^\top\widehat\bfV_v^\dml\nabla_{\bflambda} f(\widehat\bflambda_v^\dml)$ be analytical variance estimators under the same sample splitting scheme, where the explicit algebraic summations for $\widehat \bfB_v^\dml$ and $\widehat\bfSigma_v^\dml$ are provided in Section S6 of the Supplementary Materials.} To preserve the cross-fitting structure in variance estimation, the estimated H\'ajek projections used to construct $\widehat\bfSigma_v^\dml$ evaluate each pairwise score using the nuisance estimate $\widehat\zeta_p$ corresponding to the validation block $\calI_p^\pr$ containing that pair. In practice, data-adaptive estimation of $\zeta_a$ proceeds from the same pairwise specification used in Section \ref{subsec:mr-estimator}, replacing the linear index by an unspecified function: $\bbE\{w(\bfY_{ij},\bfY_{kl})|\bfU_{ij},\bfU_{kl}\}=g^{-1}\{\mu_a(\bfU_{ij},\bfU_{kl})\}$, with $\mu_a$ estimated by machine learners such as the \texttt{SuperLearner} with regression tree and neural network libraries \citep{van2007super} using input $(\bfU_{ij}^\top, \bfU_{kl}^\top)^\top$ and outcome $w(\bfY_{ij},\bfY_{kl})$. Averaging the fitted values within the pair gives $\widehat\zeta_{ik,a}=(N_iN_k)^{-1}\sum_{j=1}^{N_i}\sum_{l=1}^{N_k}g^{-1}\{\widehat\mu_a(\bfU_{ij},\bfU_{kl})\}$, the machine learning extension of Example \ref{ex:PIM}. We summarize the properties of the DML estimators below.

\begin{theorem} \label{thm:dml}
    Under Assumptions \ref{asp:sutva}--\ref{asp:cl-ran}, for $v=C,I$, if regularity conditions (C1) and (C2) in Section S6 of the Supplementary Materials hold, then $\widehat\bflambda_v^\dml\overset{p}{\to}\bflambda_v$, $m^{1/2}(\widehat\bflambda_v^\dml-\bflambda_v)\overset{d}{\to}\calN(\bm 0,\bfV_v^\dml)$ and $\widehat \bfV_v^\dml\overset{p}{\to}\bfV_v^\dml$, where $\bfV_v^\dml$ is the semiparametric efficiency bound for $\bflambda_v$.
\end{theorem}

{\color{black}Theorem \ref{thm:dml} establishes consistency, asymptotic normality, and semiparametric efficiency of the DML estimators under the regularity conditions detailed in Section S6 of the Supplementary Materials. Condition (C1) requires at least three folds and a cross-fitted nuisance estimator that is consistent in $\calL_2(\bbP)$, and condition (C2) requires a finite second moment of the contrast. These results accommodate cross-fitting in both point and variance estimation without Donsker conditions on the function classes indexed by the nuisance functions. This extends previous U-statistics DML analyses \citep{escanciano2023machine, chen2024principal}, which either did not incorporate sample splitting into variance estimation or relied on the nonparametric bootstrap. For the summary estimator $\widehat\Lambda_v^\dml$, the delta method yields $m^{1/2}(\widehat V_v^\dml)^{-1/2}(\widehat\Lambda_v^\dml-\Lambda_v)\overset{d}{\to}\calN(0,1)$ for $v=C,I$, where $\widehat V_v^\dml$ is consistent for the efficiency bound for $\Lambda_v$.

\label{par:dml-decomp}To further clarify the role of nuisance estimation, we decompose the cross-fitted U-estimating equation evaluated at the true parameter $\bflambda_v$ as
\begin{align*} \label{eq:dml-decomp}
    \binom{m}{2}^{-1} \sum_{p=1}^{|\frakI^{\pr}|} \sum_{(i,k) \in \calI_p^{\pr}} \bfpsi_v^{\eff}(\calO_i, \calO_k; \widehat{\zeta}_{p}, \bflambda_v)
    = \binom{m}{2}^{-1} \sum_{1 \le i < k \le m} \bfpsi_v^{\eff}(\calO_i, \calO_k; \zeta^0, \bflambda_v)
    + \pmb\calD_m + \pmb\calE_m,
\end{align*}
where 
$\pmb\calD_m=\binom{m}{2}^{-1}\sum_{p}|\calI_p^\pr|\,\bbE\{\bfpsi_v^{\eff}(\calO_i, \calO_k; \zeta, \bflambda_v)\}|_{\zeta=\widehat\zeta_p}$ is the population bias induced by nuisance estimation, and $\pmb\calE_m$ is the corresponding centered empirical remainder. These expectations are over independent evaluation clusters with the fitted nuisance functions held fixed. Two properties of the proposed estimating function imply that $\pmb\calD_m=\bm 0$. First, with $\kappa_{ik,a}=\bbone(A_i=a)\bbone(A_k=1-a)/\{\pi(1-\pi)\}$, the nuisance function enters $\psi_{C,a}^\eff$ only through $2^{-1}\{(1-\kappa_{ik,a})\zeta_{ik,a}+(1-\kappa_{ki,a})\zeta_{ki,a}\}$. Second, cluster randomization gives $\bbE(\kappa_{ik,a}|\mathbf W_i,\mathbf W_k)=1$. Hence, for every function $\zeta$ of $(\mathbf W_i,\mathbf W_k)$,
\begin{align*}
    &\bbE\left\{\psi_{C,a}^\eff(\calO_i,\calO_k;\zeta,\lambda_a)-\psi_{C,a}^\eff(\calO_i,\calO_k;\zeta^0,\lambda_a)\right\} \displaybreak[0]\\
    &=2^{-1}\bbE\left[\{1-\bbE(\kappa_{ik,a}|\mathbf W_i,\mathbf W_k)\}(\zeta_{ik,a}-\zeta_{ik,a}^0)+\{1-\bbE(\kappa_{ki,a}|\mathbf W_i,\mathbf W_k)\}(\zeta_{ki,a}-\zeta_{ki,a}^0)\right]=0,
\end{align*}
and the same holds for $\bfpsi_I^\eff=N_iN_k\bfpsi_C^\eff$. Equation (S22) in Section S6 of the Supplementary Materials gives the conditional version used in the proof of Theorem \ref{thm:dml}. Thus, $\bbE\{\bfpsi_v^\eff(\calO_i,\calO_k;\zeta,\bflambda_v)\}=\bm 0$ for every $\zeta$, so the population bias induced by nuisance estimation is exactly zero, rather than having a zero first-order derivative only at $\zeta^0$. This property provides the population-level explanation for the model-robust property established in Theorem \ref{thm:mr}.

Next, it remains to control the centered empirical remainder $\pmb\calE_m$. Cross-fitting makes the $p$th block of $\pmb\calE_m$ conditionally mean-zero given the clusters used to fit $\widehat\zeta_p$. Its conditional variance is at most a constant multiple of $m^{-1}\Vert\widehat\zeta_p-\zeta^0\Vert_{\calL_2}^2$. Since the number of blocks is fixed, $m^{1/2}\pmb\calE_m=\bfO_\bbP(\max_p\Vert\widehat\zeta_p-\zeta^0\Vert_{\calL_2})=\bfo_\bbP(1)$ under the consistency condition (C1). Together with $\pmb\calD_m=\bm 0$, this implies that nuisance estimation has no first-order effect on the estimator's asymptotic distribution, so Theorem \ref{thm:dml} requires no convergence rate for $\widehat\zeta_p$. In the usual DML estimation of the average treatment effect, the bias from nuisance estimation is bounded by the product of the errors of the outcome regression and the propensity score. Rate conditions such as $o_\bbP(m^{-1/4})$ make this product $o_\bbP(m^{-1/2})$ \citep[Assumption 5.1]{Chernozhukov2018}. Here $\pi$ is known, so the product is zero, as \citet[Remark 5.2]{Chernozhukov2018} note for independent data.

\label{par:pairwise}A practical challenge can occur when only a limited number of clusters are available. Under the proposed pairwise regression specification, $\zeta_{ik,a}$ is estimated by averaging predictions based on the combined covariate vector $(\bfU_{ij}^\top,\bfU_{kl}^\top)^\top$, of dimension $2p_u$. Although there are $\binom{m}{2}$ cross-cluster pairs, these pairs share clusters and therefore do not constitute independent training observations. For a nondegenerate order-two U-statistic, Hoeffding's decomposition gives a leading variance term of order $m^{-1}$ and a degenerate component of order $m^{-2}$, illustrating why the number of pairs does not imply a quadratic increase in effective sample size. Pairwise regression hence increases the covariate dimension without a corresponding increase in independent information, motivating parsimonious covariate representations, including prespecified cluster-level summaries. We illustrate this in Section \ref{sec:sim}, where, at $m=30$, the finite-sample limitations primarily concern efficiency and the accuracy of variance estimation, while the point estimates remain nearly unbiased.}

\subsection{Computationally efficient estimation using subsamples} \label{subsec:incomplete}

Since the number of pairs grows at a rate of $m^2$ and may exceed device memory for larger samples, we further propose an estimation procedure that uses subsamples to alleviate the computational resource constraint, based on incomplete U-statistics \citep{blom1976some}. Specifically, we divide the complete sample at random into $R$ disjoint subsamples of sizes $m_r$, $r=1,\ldots,R$, that are equal up to rounding. Then, on each subsample, we separately conduct the desired estimating procedure to obtain corresponding estimators, $(\widehat\bflambda_v^{\est,(r)}, \widehat\bfV_v^{\est,(r)})$ and $(\widehat\Lambda_v^{\est,(r)}, \widehat V_v^{\est,(r)})$, for $\est=\np,\mr,\dml$ and $v=C,I$. Note that for the DML estimator, each subsample is treated as if it were the complete sample, and the cross-fitting procedure is performed by forming partitions within that subsample. The final estimators are $\widehat\bflambda_v^{\est,\ast}=R^{-1}\sum_{r=1}^R\widehat\bflambda_v^{\est,(r)}$, $\widehat\bfV_v^{\est,\ast}=R^{-1}\sum_{r=1}^R\widehat\bfV_v^{\est,(r)}$, $\widehat\Lambda_v^{\est,\ast}=R^{-1}\sum_{r=1}^R\widehat\Lambda_v^{\est,(r)}$, and $\widehat V_v^{\est,\ast}=R^{-1}\sum_{r=1}^R\widehat V_v^{\est,(r)}$. The following theorem summarizes the properties of these subsample-based estimators.

\begin{theorem} \label{thm:incomplete}
    Suppose $R$ is fixed. Under the conditions of Theorem \ref{thm:np-est}, \ref{thm:mr} or \ref{thm:dml} for $\est=\np,\mr,\dml$, respectively, and for $v=C,I$, $m^{1/2}(\widehat\bflambda_v^{\est,*}-\bflambda_v)\overset{d}{\to}\calN(\bm 0,\bfV_v^\est)$ and $\widehat\bfV_v^{\est,\ast}\overset{p}{\to}\bfV_v^\est$. Moreover, $m^{1/2}(\widehat V_v^{\est,\ast})^{-1/2}(\widehat\Lambda_v^{\est,\ast}-\Lambda_v)\overset{d}{\to}\calN(0,1)$.
\end{theorem}

Section S8 of the Supplementary Materials proves Theorem \ref{thm:incomplete}. The theorem states that the subsample estimators have the same limiting distribution as the full-sample estimators, and that their variance estimators remain consistent. The subsamples can therefore be processed one at a time within device memory and then combined. In particular, the subsamples use every cluster but only about a fraction $R^{-1}$ of the cluster pairs, which loses nothing in the limit. A U-statistic behaves asymptotically like an average of terms that each involve a single cluster. The pairs that cross subsamples affect only a remainder of order $m^{-1}$, which is negligible next to the leading term of order $m^{-1/2}$. Section \ref{subsec:sim-compute} shows how much precision the subsample estimator loses in finite samples.

\section{Simulation studies} \label{sec:sim}

{\color{black}
We compare four estimators of $\lambda_{C,1}$ and $\lambda_{I,1}$. These include the NP, MR, and DML estimators in Section \ref{sec:est}, as well as the estimator of \citet{Smith2025} (denoted SJZ), which uses no covariates and does not account for informative cluster size. The MR estimator uses the PIM in Example \ref{ex:PIM} with the identity link as its working model. The DML estimator regresses the average contrast of each cluster pair on the sizes, cluster-level covariates, and covariate means of both clusters by random forests (\texttt{SL.ranger} in \texttt{SuperLearner}), with five-fold cross-fitting at the cluster level. It reports the median over five random fold assignments, with the variance aggregated as in \citet[Section 3.4]{Chernozhukov2018}. Treatment is assigned with $\pi=0.5$, and each configuration uses $500$ replications. We report the bias, the empirical standard error (ESE), the average estimated standard error (ASE), and the empirical coverage probability (ECP) of the 95\% Wald interval. Unless stated otherwise, the ASE is the average of the sandwich standard errors. With $500$ replications, the Monte Carlo standard error of a coverage near $95\%$ is about $1.0$ percentage point. Section S7.3 of the Supplementary Materials reports a third design, which varies the strength of informative cluster size.

\subsection{Simulation design 1: synthetic non-prioritized outcomes} \label{subsec:design1}

We first compare the four estimators when two non-prioritized outcomes depend nonlinearly on the baseline covariates and cluster size is informative. Each trial has $m\in\{30,60\}$ clusters, with sizes drawn uniformly from $\{2,\ldots,10\}$. Two cluster-level covariates and two individual-level covariates are generated given $N_i$. Each individual has an ordinal outcome $Y_1\in\{1,2,3\}$ and a continuous outcome $Y_2$, and both depend on the covariates through nonlinear terms. The contrast function of interest is set as $w(\bfu,\bfv)=2^{-1}\sum_{q=1}^2\{\bbone(u_q>v_q)+0.5\bbone(u_q=v_q)\}$, which gives the true estimands $\lambda_{C,1}=0.621$ and $\lambda_{I,1}=0.647$. Section S7.1 of the Supplementary Materials presents the full data-generating process. The upper panel of Web Table S2 reports the bias, ESE, ASE, and ECP.

The results empirically support Theorems \ref{thm:np-est}, \ref{thm:mr} and \ref{thm:dml}. The three proposed estimators have bias at most $0.003$, and covariate adjustment reduces their variance, as \eqref{eq:effgain-C} and \eqref{eq:effgain-I} predict. The SJZ estimator has bias $0.012$ for $\lambda_{C,1}$ and $-0.014$ for $\lambda_{I,1}$ at both sample sizes, because its target differs from both estimands when cluster size is informative. DML has the smallest ESE in every configuration. At $m=60$, its ESE for $\lambda_{C,1}$ is $0.017$, compared with $0.019$ for MR and $0.025$ for NP. The DML interval has coverage of $91.8\%$ and $92.6\%$ at $m=30$, and reaches the nominal level at $m=60$, with $94.4\%$ and $94.8\%$. The shortfall at $m=30$ arises because each random forest is trained on fewer than $30$ clusters, and its estimation error enters the estimated H\'ajek projections that form $\widehat\bfSigma_v^\dml$. The MR interval falls short, with coverage $87.2\%$ and $85.0\%$ at $m=30$ and $92.4\%$ and $93.2\%$ at $m=60$. The NP interval has coverage $91.8\%$ and $93.0\%$ at $m=30$ and $93.8\%$ and $95.6\%$ at $m=60$.

\subsection{Simulation design 2: synthetic trial resembling PPACT} \label{subsec:design2}

We next compare the four estimators in a setting that resembles the PPACT trial analyzed in Section \ref{sec:data}. Each trial has $m\in\{50,100\}$ clusters, with sizes drawn from a fixed distribution on $\{1,\ldots,9\}$, the range of cluster sizes in the trial. Each individual has four outcomes, including a binomial score $Y_1$ on $\{0,\ldots,10\}$ for pain impact, a continuous score $Y_2$ on $[0,1]$ for disability, and two satisfaction scores $Y_3$ and $Y_4$ on $\{0,\ldots,5\}$. The treatment effect on $Y_1$ grows with $N_i$, so cluster size is informative, and the covariates enter through nonlinear terms. Pairs are compared on $Y_1$ with margin $1$. Pairs tied on $Y_1$ are compared on $Y_2$ with margin $0.05$, and pairs tied on both are compared on $Y_3$ and $Y_4$ with equal weights. This rule gives $\lambda_{C,1}=0.687$ and $\lambda_{I,1}=0.697$. Section S7.1 gives the full data-generating process. The lower panel of Web Table S2 reports the bias, ESE, ASE, and ECP.

All four estimators have small bias in this design mimicking PPACT, and the largest, $-0.011$, is that of SJZ for $\lambda_{I,1}$ at $m=50$. DML is the most efficient by a wide margin. Its ESE for $\lambda_{C,1}$ is $0.051$ at $m=50$ and $0.029$ at $m=100$, compared with $0.069$ and $0.046$ for NP. The PIM working model of MR does not capture the nonlinearity in the data-generating process. MR is then less efficient than NP, with ESE $0.075$ versus $0.069$ at $m=50$, but it remains nearly unbiased, as Theorem \ref{thm:mr} predicts. The DML interval has coverage between $95.0\%$ and $96.6\%$ at both sample sizes. The MR interval again falls short, with coverage of $88.0\%$ and $87.2\%$ at $m=50$ and $94.2\%$ and $93.6\%$ at $m=100$. The NP interval has coverage between $92.8\%$ and $96.2\%$.

\subsection{Improving variance estimation for the MR estimator} \label{subsec:sim_variance}

Simulation designs 1 and 2 show that the sandwich interval of the MR estimator under-covers in finite samples. The working model parameter $\widehat\bfvartheta_a$ is fitted on the same clusters at which the estimating equation \eqref{eq:dr-equations} is evaluated, and hence the fitted residuals $\overline w(\mathbf Y_i,\mathbf Y_k)-\zeta_{ik,a}(\widehat\bfvartheta_a)$ can often be biased toward zero. The leave-one-cluster-out jackknife variance estimator can be an alternative that addresses this downward bias. Let $\widehat\bflambda_v^{\mr,(-i)}$ be the MR estimator computed without cluster $i$, with $\bfvartheta_a$ refitted, and let $\overline\bflambda_v^\mr=m^{-1}\sum_{i=1}^m\widehat\bflambda_v^{\mr,(-i)}$. The jackknife estimator of the asymptotic covariance matrix of $m^{1/2}\widehat\bflambda_v^\mr$ is $\widehat\bfV_{v,\mathrm{jack}}^\mr=(m-1)\sum_{i=1}^m\left(\widehat\bflambda_v^{\mr,(-i)}-\overline\bflambda_v^\mr\right)\left(\widehat\bflambda_v^{\mr,(-i)}-\overline\bflambda_v^\mr\right)^\top$. 
Because the clusters are independent and identically distributed under Assumption \ref{asp:super-pop}, the jackknife variance estimate is biased upward for the variance based on $m-1$ clusters \citep{efron1981jackknife}, so the interval tends to be conservative. We therefore use the same jackknife for the NP estimator, whose sandwich interval also falls short with $50$ or fewer clusters. Of note, we do not use it for DML, since cross-fitting already evaluates each pairwise score with a nuisance estimate fitted on other clusters, and the jackknife would require $m$ further rounds of cross-fitting.

We conduct a complementary study to compare the sandwich and jackknife standard errors of the MR and NP estimators in Designs 1 and 2. Web Table S3 reports the ESE, the average sandwich and jackknife standard errors, and the coverage of the two 95\% intervals. The jackknife largely corrects the shortfall in sandwich interval coverage. In simulation design 1 at $m=30$, the average sandwich standard error of MR for $\lambda_{I,1}$ is $0.021$ against an ESE of $0.027$, while the average jackknife standard error is $0.032$. The coverage rises from $85.0\%$ to $95.2\%$. The jackknife gives MR coverage between $95.2\%$ and $97.0\%$ in every configuration, and the upper end reflects its upward bias. For NP, it raises the coverage at $m=30$ from $91.8\%$ and $93.0\%$ to $93.6\%$ and $94.2\%$, and gives coverage between $93.6\%$ and $96.2\%$ in every configuration. 

\subsection{Illustration of computation savings with subsamples} \label{subsec:sim-compute}

An additional simulation is conducted to illustrate the computing time saved by the subsample estimator in Section \ref{subsec:incomplete} and the precision it concedes. To focus ideas, we simulate trials with $m=90$ clusters from simulation design 2. We compare DML on the full sample, MR on the full sample, and MR computed on $R=3$ disjoint random subsamples of $30$ clusters and averaged as in Section \ref{subsec:incomplete}. DML uses the sandwich standard error, and both MR estimators use the jackknife standard error in Section \ref{subsec:sim_variance}. Web Table S5 reports the computing time, bias, ESE, ASE, and ECP. The simulation results empirically support Theorem \ref{thm:incomplete}. All three estimators are nearly unbiased, with bias at most $0.002$, and subsampling cuts the computing time by an order of magnitude at a moderate loss of precision in finite samples. On an Apple M1 processor, full DML takes $128.87$ seconds per trial, and full MR takes $28.66$ seconds. The subsample MR estimator takes $3.00$ seconds, which is $43$ times faster than DML and $9.6$ times faster than full MR. Its ESE for $\lambda_{C,1}$ is $0.060$, compared with $0.052$ for full MR and $0.032$ for DML. The loss comes from the discarded cross-subsample pairs and from fitting the working model on $30$ clusters at a time. The DML interval has coverage $95.4\%$ and $96.0\%$, and the full MR interval has coverage $95.0\%$ and $95.6\%$. The subsample MR interval has coverage $97.2\%$ and $97.0\%$, because each working model in the jackknife is fitted on fewer than $30$ clusters. Section S8.3 of the Supplementary Materials turns these results into an implementation guide.
}

\section{Application to the PPACT cluster-randomized trial} \label{sec:data}

{\color{black}
We apply our methods to the PPACT trial \citep{DeBar2022}. PPACT was a CRT conducted in primary care in several Kaiser Permanente regions. It compared usual care with an interdisciplinary care model that placed behavioral health specialists, nurse care managers, physical therapists, and pharmacists in primary care. The participating patients had chronic pain and were on long-term opioid therapy. Each randomized cluster was a clinic-based group of 1 to 6 providers and their patients. We used four patient-reported outcomes. The primary outcome $Y_1$ is pain impact at 12 months, measured on the four-item PEGS scale from 0 to 10. The outcome $Y_2$ is pain-related disability at 12 months, measured by the 24-item Roland--Morris Disability Questionnaire (RMDQ) on a range from 0 to 1. The outcomes $Y_3$ and $Y_4$ are satisfaction with primary care and with pain services at 6 months, each on a six-point scale from 0 (very dissatisfied) to 5 (very satisfied). Lower values of $Y_1$ and $Y_2$ and higher values of $Y_3$ and $Y_4$ are considered more favorable. Because the four outcomes are on different scales, a mean contrast across them is not clinically meaningful. The analysis includes the 560 patients with all four outcomes, in 104 clusters of 1 to 9 patients (mean $5.385$). A causal reading of this complete-case analysis requires assumptions on outcome observation.

In each analysis, we estimate the cluster-pair and individual-pair win ratios $\Lambda_v=\lambda_{v,1}/\lambda_{v,0}$, $v=C,I$, of Remark \ref{rmk:estimand}. With half credit for ties, each tie counts as half a win for each arm, as in the win odds of Remark \ref{rmk:estimand}. A ratio above one favors the interdisciplinary care model. We compute the NP, MR, and DML estimators, with the MR and DML nuisance models specified as in Section \ref{sec:sim}. The adjusted estimators use the following covariates, including gender, age, a diagnosis of anxiety or depression, the number of pain types, the average morphine milligram equivalent dose, benzodiazepine dispensing, baseline PEGS, baseline satisfaction with primary care, and an indicator of two or more chronic conditions among diabetes, cardiovascular disorders, hypertension, and chronic pulmonary disease. Simulation design 2 in Section \ref{subsec:design2} shows that with $100$ clusters the sandwich interval of NP has coverage close to $95\%$, while that of MR still falls short. We therefore use sandwich standard errors for NP and DML and jackknife standard errors for MR.

\subsection{Analysis of prioritized outcomes} \label{subsec:data-analysis-prioritized}

We first rank the outcomes with the priority hierarchy of $Y_1$ first, $Y_2$ second, and $Y_3$ and $Y_4$ jointly third. A pair is first compared on PEGS. The treated member wins if its PEGS is more than one point lower, loses if it is more than one point higher, and ties otherwise. The 1-point margin is the minimally important difference that \citet{Reed2024} report for the PEG, which omits the sleep item of PEGS. A pair tied on PEGS is compared on RMDQ in the same way with margin $0.05$, or $1.2$ of 24 points, below the 5-point minimal important change of \citet{Ostelo2008}. A pair tied on both is compared on $Y_3$ and $Y_4$ with the contrast $\bbone(u>v)+0.5\bbone(u=v)$ of \citet{brunner2021win}, each with weight $0.5$. Among all pairs, $28.7\%$ are tied on $Y_1$, $15.1\%$ are tied on $Y_2$, and $6.1\%$ are tied on both. We consider three outcome sets, namely $Y_1$ alone, $(Y_1,Y_2)$ and $(Y_1,\ldots,Y_4)$. Table \ref{tab:data-results-p} reports the estimates, standard errors, and 95\% confidence intervals.
}

\begin{table}[t]
    \caption{\color{black}Win ratio estimates in the PPACT trial with prioritized outcomes. NP: nonparametric; MR: model-robust; DML: debiased machine learning. EST: estimate; SE: standard error; CI: 95\% confidence interval. $^{\ast}$: the 95\% confidence interval does not contain 1.} \label{tab:data-results-p}
    \centering
    {\color{black}\footnotesize\setlength{\tabcolsep}{2pt}
    \begin{tabular}{cc ccc ccc}
        \toprule
        & & \multicolumn{3}{c}{$\Lambda_C$} & \multicolumn{3}{c}{$\Lambda_I$} \\
        \cmidrule(lr){3-5} \cmidrule(lr){6-8}
        Outcomes &  & NP & MR & DML & NP & MR & DML \\
        \midrule
        \multirow{2}{*}{$Y_1$} & EST (SE) & 1.349 (.209) & 1.258 (.157) & 1.246 (.149) & 1.393$^{\ast}$ (.192) & 1.240 (.144) & 1.244 (.138) \\
        & CI & (.940,1.758) & (.952,1.565) & (.955,1.538) & (1.016,1.770) & (.958,1.522) & (.973,1.515) \\
        \cmidrule{1-8}
        \multirow{2}{*}{$Y_1,Y_2$} & EST (SE) & 1.298 (.161) & 1.227 (.126) & 1.236$^{\ast}$ (.116) & 1.375$^{\ast}$ (.151) & 1.253$^{\ast}$ (.114) & 1.271$^{\ast}$ (.108) \\
        & CI & (.981,1.614) & (.980,1.474) & (1.009,1.463) & (1.080,1.671) & (1.029,1.476) & (1.059,1.482) \\
        \cmidrule{1-8}
        \multirow{2}{*}{$Y_1,\ldots,Y_4$} & EST (SE) & 1.293 (.151) & 1.225 (.118) & 1.227$^{\ast}$ (.108) & 1.363$^{\ast}$ (.142) & 1.248$^{\ast}$ (.107) & 1.256$^{\ast}$ (.101) \\
        & CI & (.998,1.588) & (.994,1.456) & (1.015,1.440) & (1.085,1.640) & (1.038,1.459) & (1.058,1.455) \\
        \bottomrule
    \end{tabular}
    }
\end{table}

{\color{black}
All eighteen estimates exceed one, so every analysis points in favor of the interdisciplinary care model. The MR and DML estimates are smaller than the NP estimates, by $0.06$ to $0.15$. Relative to NP, MR reduces the standard error by $22\%$ to $25\%$ and DML by $28\%$ to $29\%$. Adding outcomes to the hierarchy also reduces the standard errors, because comparisons tied on PEGS are then decided by the lower-priority outcomes. For DML and $\Lambda_C$, the standard error falls from $0.149$ with $Y_1$ alone to $0.116$ with $(Y_1,Y_2)$ and $0.108$ with all four outcomes. With $Y_1$ alone, only the NP interval for $\Lambda_I$ excludes one. Once $Y_2$ enters, the DML intervals exclude one for both estimands. The MR intervals then exclude one for $\Lambda_I$ but not for $\Lambda_C$, and the NP intervals exclude one for $\Lambda_I$ only.

\subsection{Analysis of non-prioritized outcomes} \label{subsec:data-analysis-non-prioritized}

We next consider a setting where the outcomes are non-prioritized. Each outcome is compared with its rule from Section \ref{subsec:data-analysis-prioritized}, and the overall contrast is the equally weighted average of the outcome-specific contrasts. We consider two outcome sets, $(Y_1,Y_2)$ and $(Y_1,\ldots,Y_4)$. With $Y_1$ alone, the analysis coincides with that of Section \ref{subsec:data-analysis-prioritized}. Table \ref{tab:data-results-np} reports the estimates, standard errors, and 95\% confidence intervals.
}

\begin{table}[htbp]
    \caption{\color{black}Win ratio estimates in the PPACT trial with non-prioritized outcomes. Notation as in Table \ref{tab:data-results-p}.} \label{tab:data-results-np}
    \centering
    {\color{black}\footnotesize\setlength{\tabcolsep}{2pt}
    \begin{tabular}{cc ccc ccc}
        \toprule
        & & \multicolumn{3}{c}{$\Lambda_C$} & \multicolumn{3}{c}{$\Lambda_I$} \\
        \cmidrule(lr){3-5} \cmidrule(lr){6-8}
        Outcomes &  & NP & MR & DML & NP & MR & DML \\
        \midrule
        \multirow{2}{*}{$Y_1,Y_2$} & EST (SE) & 1.349 (.179) & 1.260 (.143) & 1.268$^{\ast}$ (.133) & 1.450$^{\ast}$ (.167) & 1.301$^{\ast}$ (.128) & 1.321$^{\ast}$ (.119) \\
        & CI & (.997,1.700) & (.978,1.541) & (1.009,1.528) & (1.122,1.777) & (1.051,1.552) & (1.089,1.553) \\
        \cmidrule{1-8}
        \multirow{2}{*}{$Y_1,\ldots,Y_4$} & EST (SE) & 1.336$^{\ast}$ (.107) & 1.281$^{\ast}$ (.096) & 1.288$^{\ast}$ (.086) & 1.351$^{\ast}$ (.100) & 1.287$^{\ast}$ (.080) & 1.291$^{\ast}$ (.076) \\
        & CI & (1.126,1.546) & (1.093,1.469) & (1.119,1.457) & (1.155,1.548) & (1.130,1.444) & (1.142,1.440) \\
        \bottomrule
    \end{tabular}
    }
\end{table}

{\color{black}
All twelve estimates exceed one. With $(Y_1,Y_2)$, the intervals for $\Lambda_I$ exclude one across all three estimators. For $\Lambda_C$ only the DML interval does, with estimate $1.268$ and interval $(1.009, 1.528)$. With all four outcomes, every interval excludes one. DML again has the smallest standard errors. For every estimator and outcome set, $\widehat\Lambda_I$ has a larger magnitude than $\widehat\Lambda_C$, and the gap is largest for NP with $(Y_1,Y_2)$, at $1.450$ against $1.349$. The two estimands coincide when cluster size is not informative (Section \ref{sec:setting}). The consistent gap between $\widehat\Lambda_I$ and $\widehat\Lambda_C$ suggests potentially informative cluster size in PPACT.

\subsection{Sensitivity to the equivalence margins} \label{subsec:sensitivity}

We perform an additional sensitivity analysis to examine how the prioritized win ratios depend on the equivalence margins, which are part of the definition of $w$ and hence of the estimand. The primary margins of 1 point on PEGS and $0.05$ on RMDQ were fixed before the analysis. We recompute the DML estimates for $(Y_1,\ldots,Y_4)$ under two alternatives: (i) strict margins of 2 points on PEGS and $0.10$ on RMDQ, which leads to more ties and passes more comparisons down the hierarchy; (ii) lenient margins of zero, where any difference on both decides a comparison, with only exact ties on PEGS reaching the lower-priority outcomes. Table \ref{tab:sensitivity} reports the estimates, standard errors, and 95\% confidence intervals under the three specifications.
}

\begin{table}[htbp]
    \caption{\color{black}DML win ratio estimates in PPACT with prioritized outcomes $(Y_1,\ldots,Y_4)$ under different equivalence margins. Lenient: 0 on PEGS and 0 on RMDQ; Primary: 1 on PEGS and $0.05$ on RMDQ; Strict: 2 on PEGS and $0.10$ on RMDQ. EST: estimate; SE: standard error; CI: 95\% confidence interval. $^{\ast}$: the 95\% confidence interval does not contain 1.}
    \label{tab:sensitivity}
    \centering
    {\color{black}
    \begin{tabular}{cc ccc}
        \toprule
        & & \multicolumn{3}{c}{Margins} \\
        \cmidrule(lr){3-5}
        Estimand &  & \multicolumn{1}{c}{Lenient} & \multicolumn{1}{c}{Primary} & \multicolumn{1}{c}{Strict} \\
        \cmidrule{1-5}
        \multirow{2}{*}{$\Lambda_C$} & EST (SE) & 1.176 (.106) & 1.227$^{\ast}$ (.108) & 1.279$^{\ast}$ (.116) \\
        & CI & (.968,1.383) & (1.015,1.440) & (1.051,1.507) \\
        \cmidrule{1-5}
        \multirow{2}{*}{$\Lambda_I$} & EST (SE) & 1.180 (.098) & 1.256$^{\ast}$ (.101) & 1.329$^{\ast}$ (.105) \\
        & CI & (.988,1.372) & (1.058,1.455) & (1.124,1.534) \\
        \bottomrule
    \end{tabular}
    }
\end{table}

{\color{black}
All six estimates exceed one, and interestingly, their magnitudes increase with the margins. The estimate of $\Lambda_I$ is $1.180$ under lenient margins, $1.256$ under the primary margins, and $1.329$ under strict margins, and $\Lambda_C$ moves in the same direction. The intervals exclude one under the primary and strict margins but not under the lenient margins. With zero margins, a difference of a fraction of a point in PEGS counts as a win or a loss, and the disability and satisfaction outcomes enter only through exact ties. The lenient equivalence margin therefore studies whether the intervention improves pain impact by any amount, including differences below the minimal clinically important difference. The primary equivalence margin, on the other hand, studies whether the intervention improves pain impact by a clinically important amount and, for pairs tied on pain impact, whether it improves disability or satisfaction. These analyses are not necessarily contradictory. They target distinct estimands that address different scientific questions. The observed sensitivity to the margins underscores the importance of prespecifying clinically justified margins that align with the scientific question of interest.}

\vspace{-0.5cm}
\section{Discussion} \label{sec:conclusion}

{\color{black}
\label{par:discussion}We proposed cluster-pair and individual-pair generalized causal effects, defined through a pairwise contrast function, for CRTs with multiple outcomes. They cover prioritized and non-prioritized outcomes and include the net benefit, win ratio, win odds, and global treatment effect as special cases. The cluster-pair estimand weights clusters equally, the individual-pair estimand weights individuals equally, and the two can differ when cluster size is informative \citep{kahan2023estimands}. We estimated both by weighted clustered U-statistics, derived their EIFs, and characterized the efficiency gain from covariate adjustment. The model-robust and DML estimators built on the EIFs remain consistent when the nuisance model is misspecified, and the DML estimator attains the efficiency bound under the conditions of Theorem \ref{thm:dml}. Furthermore, the DML estimator is therefore $m^{1/2}$-consistent and asymptotically normal as long as the nuisance estimator is consistent, with no requirement on its rate of convergence. Distinct from prior work \citep{escanciano2023machine,chen2024principal}, our cross-fitted sandwich variance estimator is also proved to be consistent without Donsker conditions. 

\label{par:perspectives}In practice, multiple outcomes can also be analyzed separately, with simultaneous confidence intervals or a joint confidence region for the outcome-specific treatment effects. Such analyses reveal which outcomes improve or worsen, but do not by themselves specify how benefits and harms across outcomes should be combined into an overall clinical assessment. A related predictive literature constructs prediction regions with joint coverage for a vector of outcomes, including the recent conformal methods of \citet{Klein2025}, \citet{Thurin2025}, and \citet{Gao2026}. Although prediction regions for outcomes serve a different purpose from confidence regions for treatment effects, both approaches preserve information across outcome dimensions. Our framework complements these existing perspectives by defining a scalar estimand through a prespecified, clinically meaningful comparison rule, particularly when the outcomes admit a clinically justified hierarchy, as is the case with recent clinical studies \citep{buyse2010generalized,Pocock2012,bebu2016large}. This rule makes explicit how outcomes are prioritized and which differences are considered clinically meaningful, thereby anchoring the overall assessment to the scientific question of interest. Combining information across outcomes may also improve power in some settings, including when individual outcomes are infrequent or separate claims require multiplicity adjustment, although such gains are not guaranteed. The main limitation is that an overall summary can mask opposing treatment effects across outcomes or obscure which outcomes drive the result. We therefore view the GCE as an initial overall assessment, potentially to be reported alongside outcome-specific effects. Used together, these analyses can provide a clinically interpretable summary of the intervention's overall effect while retaining the detail needed to understand its benefits and harms.}


\bibliographystyle{jasa3}
\bibliography{CWE}

\begin{thebibliography}{33}
\newcommand{\enquote}[1]{``#1''}
\expandafter\ifx\csname natexlab\endcsname\relax\def\natexlab#1{#1}\fi
\expandafter\ifx\csname url\endcsname\relax
  \def\url#1{{\tt #1}}\fi
\expandafter\ifx\csname urlprefix\endcsname\relax\def\urlprefix{URL }\fi

\bibitem[\protect\citeauthoryear{Balzer, Zheng, van~der Laan, and
  Petersen}{Balzer et~al.}{2019}]{balzer2019new}
Balzer, L.~B., Zheng, W., van~der Laan, M.~J., and Petersen, M.~L. (2019),
  \enquote{A new approach to hierarchical data analysis: targeted maximum
  likelihood estimation for the causal effect of a cluster-level exposure,}
  {\em Statistical Methods in Medical Research\/}, 28, 1761--1780.

\bibitem[\protect\citeauthoryear{Bebu and Lachin}{Bebu and
  Lachin}{2016}]{bebu2016large}
Bebu, I. and Lachin, J.~M. (2016), \enquote{Large sample inference for a win
  ratio analysis of a composite outcome based on prioritized components,} {\em
  Biostatistics\/}, 17, 178--187.

\bibitem[\protect\citeauthoryear{Blom}{Blom}{1976}]{blom1976some}
Blom, G. (1976), \enquote{Some properties of incomplete U-statistics,} {\em
  Biometrika\/}, 63, 573--580.

\bibitem[\protect\citeauthoryear{Brunner, Vandemeulebroecke, and
  M{\"u}tze}{Brunner et~al.}{2021}]{brunner2021win}
Brunner, E., Vandemeulebroecke, M., and M{\"u}tze, T. (2021), \enquote{Win
  odds: an adaptation of the win ratio to include ties,} {\em Statistics in
  Medicine\/}, 40, 3367--3384.

\bibitem[\protect\citeauthoryear{Bugni, Canay, Shaikh, and Tabord-Meehan}{Bugni
  et~al.}{2025}]{Bugni2024}
Bugni, F., Canay, I.~A., Shaikh, A.~M., and Tabord-Meehan, M. (2025),
  \enquote{Inference for cluster randomized experiments with nonignorable
  cluster sizes,} {\em Journal of Political Economy Microeconomics\/}, 3,
  255--288.

\bibitem[\protect\citeauthoryear{Buyse}{Buyse}{2010}]{buyse2010generalized}
Buyse, M. (2010), \enquote{Generalized pairwise comparisons of prioritized
  outcomes in the two-sample problem,} {\em Statistics in Medicine\/}, 29,
  3245--3257.

\bibitem[\protect\citeauthoryear{Chen and Li}{Chen and
  Li}{2026}]{chen2024principal}
Chen, X. and Li, F. (2026), \enquote{Principal stratification with U-statistics
  under principal ignorability,} {\em Journal of the Royal Statistical Society:
  Series B (Statistical Methodology)\/}.

\bibitem[\protect\citeauthoryear{Chernozhukov, Chetverikov, Demirer, Duflo,
  Hansen, Newey, and Robins}{Chernozhukov et~al.}{2018}]{Chernozhukov2018}
Chernozhukov, V., Chetverikov, D., Demirer, M., Duflo, E., Hansen, C., Newey,
  W., and Robins, J. (2018), \enquote{{Double/debiased machine learning for
  treatment and structural parameters},} {\em The Econometrics Journal\/}, 21,
  C1--C68.

\bibitem[\protect\citeauthoryear{DeBar, Mayhew, Benes, Bonifay, Deyo, Elder,
  Keefe, Leo, McMullen, Owen-Smith, Smith, Trinacty, and Vollmer}{DeBar
  et~al.}{2022}]{DeBar2022}
DeBar, L., Mayhew, M., Benes, L., Bonifay, A., Deyo, R.~A., Elder, C.~R.,
  Keefe, F.~J., Leo, M.~C., McMullen, C., Owen-Smith, A., Smith, D.~H.,
  Trinacty, C.~M., and Vollmer, W.~M. (2022), \enquote{A primary care-based
  cognitive behavioral therapy intervention for long-term opioid users with
  chronic pain,} {\em Annals of Internal Medicine\/}, 175, 46--55. PMID:
  34724405.

\bibitem[\protect\citeauthoryear{Efron and Stein}{Efron and
  Stein}{1981}]{efron1981jackknife}
Efron, B. and Stein, C. (1981), \enquote{The Jackknife Estimate of Variance,}
  {\em The Annals of Statistics\/}, 9, 586--596.

\bibitem[\protect\citeauthoryear{Escanciano and Terschuur}{Escanciano and
  Terschuur}{2023}]{escanciano2023machine}
Escanciano, J.~C. and Terschuur, J.~R. (2023), \enquote{{Machine learning
  inference on inequality of opportunity},} {\em arXiv preprint
  arXiv.2206.05235\/}.

\bibitem[\protect\citeauthoryear{Fang, Cao, and Li}{Fang
  et~al.}{2025{\natexlab{a}}}]{fang2025sample}
Fang, X., Cao, Z., and Li, F. (2025{\natexlab{a}}), \enquote{Sample size
  determination for win statistics in cluster-randomized trials,} {\em arXiv
  preprint arXiv:2510.22709\/}.

\bibitem[\protect\citeauthoryear{Fang, Uno, and Li}{Fang
  et~al.}{2025{\natexlab{b}}}]{fang2025while}
Fang, X., Uno, H., and Li, F. (2025{\natexlab{b}}), \enquote{While-alive
  regression analysis of composite survival endpoints,} {\em Biostatistics\/},
  26, kxaf047.

\bibitem[\protect\citeauthoryear{Gao and Han}{Gao and Han}{2026}]{Gao2026}
Gao, C. and Han, L. (2026), \enquote{{FACTOR}: Fairness-aligned conformal
  transport for optimal regions,} Statistics and Trustworthy AI for Cross
  (X)-Domain Acceleration (STAI-X) Workshop.

\bibitem[\protect\citeauthoryear{Kahan, Li, Copas, and Harhay}{Kahan
  et~al.}{2023}]{kahan2023estimands}
Kahan, B.~C., Li, F., Copas, A.~J., and Harhay, M.~O. (2023),
  \enquote{Estimands in cluster-randomized trials: choosing analyses that
  answer the right question,} {\em International Journal of Epidemiology\/},
  52, 107--118.

\bibitem[\protect\citeauthoryear{Kallus and Mao}{Kallus and
  Mao}{2025}]{Kallus2025}
Kallus, N. and Mao, X. (2025), \enquote{On the role of surrogates in the
  efficient estimation of treatment effects with limited outcome data,} {\em
  Journal of the Royal Statistical Society Series B: Statistical
  Methodology\/}, 87, 480--509.

\bibitem[\protect\citeauthoryear{Klein, Bethune, Ndiaye, and Cuturi}{Klein
  et~al.}{2025}]{Klein2025}
Klein, M., Bethune, L., Ndiaye, E., and Cuturi, M. (2025),
  \enquote{Multivariate conformal prediction using optimal transport,} {\em
  arXiv preprint arXiv:2502.03609\/}.

\bibitem[\protect\citeauthoryear{Lee and Dehling}{Lee and
  Dehling}{2005}]{Lee2005}
Lee, M.-L.~T. and Dehling, H.~G. (2005), \enquote{Generalized two-sample
  $U$-statistics for clustered data,} {\em Statistica Neerlandica\/}, 59,
  313--323.

\bibitem[\protect\citeauthoryear{Li, Tong, Fang, Cheng, Kahan, and Wang}{Li
  et~al.}{2025}]{li2025model}
Li, F., Tong, J., Fang, X., Cheng, C., Kahan, B.~C., and Wang, B. (2025),
  \enquote{Model-Robust Standardization in Cluster-Randomized Trials,} {\em
  Statistics in Medicine\/}, 44, e70270.

\bibitem[\protect\citeauthoryear{Mao}{Mao}{2017}]{Mao2018}
Mao, L. (2017), \enquote{{On causal estimation using U-statistics},} {\em
  Biometrika\/}, 105, 215--220.

\bibitem[\protect\citeauthoryear{Ostelo, Deyo, Stratford, Waddell, Croft,
  Von~Korff, Bouter, and De~Vet}{Ostelo et~al.}{2008}]{Ostelo2008}
Ostelo, R.~W., Deyo, R.~A., Stratford, P., Waddell, G., Croft, P., Von~Korff,
  M., Bouter, L.~M., and De~Vet, H.~C. (2008), \enquote{Interpreting change
  scores for pain and functional status in low back pain: towards international
  consensus regarding minimal important change,} {\em Spine\/}, 33, 90--94.

\bibitem[\protect\citeauthoryear{Pocock, Ariti, Collier, and Wang}{Pocock
  et~al.}{2011}]{Pocock2012}
Pocock, S.~J., Ariti, C.~A., Collier, T.~J., and Wang, D. (2011), \enquote{{The
  win ratio: A new approach to the analysis of composite endpoints in clinical
  trials based on clinical priorities},} {\em European Heart Journal\/}, 33,
  176--182.

\bibitem[\protect\citeauthoryear{Reed~II, Stump, Monahan, and Kroenke}{Reed~II
  et~al.}{2024}]{Reed2024}
Reed~II, D.~E., Stump, T.~E., Monahan, P.~O., and Kroenke, K. (2024),
  \enquote{Comparable minimally important differences and responsiveness of
  brief pain inventory and PEG pain scales across 6 trials,} {\em The Journal
  of Pain\/}, 25, 142--152.

\bibitem[\protect\citeauthoryear{Smith, Jairath, and Zou}{Smith
  et~al.}{2025}]{Smith2025}
Smith, E.~D., Jairath, V., and Zou, G. (2025), \enquote{Rank-based estimators
  of global treatment effects for cluster randomized trials with multiple
  endpoints on different scales,} {\em Statistical Methods in Medical
  Research\/}, 34, 1267--1289. PMID: 40368381.

\bibitem[\protect\citeauthoryear{Su and Ding}{Su and Ding}{2021}]{Su2021}
Su, F. and Ding, P. (2021), \enquote{Model-assisted analyses of
  cluster-randomized experiments,} {\em Journal of the Royal Statistical
  Society: Series B (Statistical Methodology)\/}, 83, 994--1015.

\bibitem[\protect\citeauthoryear{Thas, Neve, Clement, and Ottoy}{Thas
  et~al.}{2012}]{thas2012probabilistic}
Thas, O., Neve, J.~D., Clement, L., and Ottoy, J.-P. (2012),
  \enquote{Probabilistic index models,} {\em Journal of the Royal Statistical
  Society Series B: Statistical Methodology\/}, 74, 623--671.

\bibitem[\protect\citeauthoryear{Thurin, Nadjahi, and Boyer}{Thurin
  et~al.}{2025}]{Thurin2025}
Thurin, G., Nadjahi, K., and Boyer, C. (2025), \enquote{Optimal transport-based
  conformal prediction,} {\em arXiv preprint arXiv:2501.18991\/}.

\bibitem[\protect\citeauthoryear{Tsiatis}{Tsiatis}{2006}]{Tsiatis2006}
Tsiatis, A.~A. (2006), {\em Semiparametric Theory and Missing Data\/}, Springer
  New York, NY.

\bibitem[\protect\citeauthoryear{Turner, Prague, Gallis, Li, and Murray}{Turner
  et~al.}{2017}]{Turner2017review}
Turner, E.~L., Prague, M., Gallis, J.~A., Li, F., and Murray, D.~M. (2017),
  \enquote{Review of recent methodological developments in group-randomized
  trials: part 2—analysis,} {\em American Journal of Public Health\/}, 107,
  1078--1086.

\bibitem[\protect\citeauthoryear{van~der Laan, Polley, and Hubbard}{van~der
  Laan et~al.}{2007}]{van2007super}
van~der Laan, M.~J., Polley, E.~C., and Hubbard, A.~E. (2007), \enquote{Super
  learner,} {\em Statistical Applications in Genetics and Molecular Biology\/},
  6.

\bibitem[\protect\citeauthoryear{{van der Vaart}}{{van der
  Vaart}}{1998}]{vandervaart1998}
{van der Vaart}, A. (1998), {\em Asymptotic Statistics\/}, Cambridge, UK:
  Cambridge University Press.

\bibitem[\protect\citeauthoryear{Wang, Park, Small, and Li}{Wang
  et~al.}{2024}]{wang2024model}
Wang, B., Park, C., Small, D.~S., and Li, F. (2024), \enquote{Model-robust and
  efficient covariate adjustment for cluster-randomized experiments,} {\em
  Journal of the American Statistical Association\/}, 119, 2959--2971.

\bibitem[\protect\citeauthoryear{Wolfe and Hogg}{Wolfe and
  Hogg}{1971}]{wolfe1971constructing}
Wolfe, D.~A. and Hogg, R.~V. (1971), \enquote{On constructing statistics and
  reporting data,} {\em The American Statistician\/}, 25, 27--30.

\end{thebibliography}

\end{document}